# A Numerical Study of Thermal-Hydraulic-Mechanical (THM) Simulation with the Application of Thermal Recovery in Fractured Shale Gas Reservoirs*

HanYi Wang, The University of Texas at Austin

**Summary**

Shale gas is playing an important role in transforming global energy markets with increasing demands for cleaner energy in the future. One major difference in shale gas reservoirs is that a considerable amount of gas is adsorbed. Up to 85% of the total gas within shale may be found adsorbed on clay and kerogen. How much of the adsorbed gas can be produced has a significant impact on ultimate recovery. Even with improving fracturing and horizontal well technologies, average gas recovery factors in U.S. shale plays is only ~30% with primary depletion. Adsorbed gas can be desorbed by lowering pressure and raising temperature, reservoir flow capacity can be also influenced by temperature, so there is a big prize to be claimed using thermal stimulation techniques to enhance recovery. To date, not much work has been done on thermal stimulation of gas shale reservoirs.

In this article, we present general formulations to simulate gas production in fractured shale gas reservoirs for the first time, with fully coupled thermal-hydraulic-mechanical (THM) properties. The unified shale gas reservoir model developed in this study enable us to investigate multi-physics phenomena in shale gas formations. Thermal stimulation of fractured gas reservoirs by heating propped fractures is proposed and investigated. This study provides some fundamental insight into real gas flow in nano-pore space and gas adsorption/desorption behavior in fractured gas shales under various in-situ conditions and sets a foundation for future research efforts in the area of enhanced recovery of shale gas reservoirs.

We find that thermal stimulation of shale gas reservoirs has the potential to enhance recovery significantly by enhancing the overall flow capacity and releasing adsorbed gas that cannot be recovered by depletion, but the process may be hampered by the low rate of purely conductive heat transfer, if only the surfaces of hydraulic fractures are heated.

## Introduction

Unconventional resources—shale gas and liquids, coalbed methane, tight gas and heavy oil—are called "unconventionals" because in order to economically access and produce these hydrocarbons, unconventional methods and expertise are required. Better reservoir knowledge and increasingly sophisticated technologies (especially horizontal drilling and hydraulic fracturing) make the production of unconventional resources economically viable and more efficient. This efficiency is bringing shale reservoirs, tight gas and oil, and coalbed methane into the reach of more companies around the world. With ever increasing demanding for cleaner energy, unconventional gas reservoirs are expected to play a vital role in satisfying the global needs for gas in the future. The major component of unconventional gas reservoirs comprises of shale gas. Shales and silts are the most abundant sedimentary rocks in the earth's crust and it is evident from the recent year's activities in shale gas plays that in the future shale gas will constitute the largest component in gas production globally. According to GSGI, there are more than 688 shales worldwide in 142 basins and 48 major shale basins are located in 32 countries (Newell, 2011). Shale gas exploitation is no longer an uneconomic venture with the availability of improved technology, as the demand and preference for cleaner form of hydrocarbon are in ever greater demands, especially in country like China, where the main energy resource still comes from coal. Unlike conventional gas reservoirs, shale gas reservoirs have very low permeability and are economical only when hydraulically fractured. The key techniques that allow extracting shale gas commercially such as horizontal drilling and hydraulic fracturing, are expected to improve with time; however as better stimulation techniques are becoming attainable, it is important to have a better understanding of shale gas reservoir behavior in order to apply these techniques in an efficient fashion. One important aspect of shale gas reservoirs which needs special consideration is the adsorption/desorption phenomenon.

In organic porous media, gas can be stored as compressed fluid inside the pores or it can be adsorbed by the solid matrix. Similar to surface tension, adsorption is a consequence of surface energy (Gregg and Sing, 1982), which causes gas molecules to get bonded to the surface of the rock grains. The gas adsorption in the shale-gas system is primarily controlled by the presence of organic matter and the gas adsorption capacity, which depends on TOC (Total Organic Carbon), organic matter type, thermal maturity and clay minerals (Ambrose et al., 2010; Passey et al., 2010). Generally, the higher the TOC content, the greater the gas adsorption capacity. In addition, a large number of nanopores lead to significant nanoporosity in shale formations, which increases the gas adsorption surface area substantially. The amount of adsorbed gas varies from 35-58%

---

(Barnett Shale, USA) up to 60-85% (Lewis Shale, USA) of total gas initial in-place (Darishchev et al., 2013). Presently, the only method for accurately determining the adsorbed gas in a formation is through core sampling and analysis. However, understanding the effects that initial adsorption, and moreover, desorption has on gas production will increase the effectiveness of reservoir management in these challenging environments.

Besides horizontal drilling and hydraulic fracturing, which ignited the momentum of "shale revolution", thermal stimulation methods have been widely used for unconventional reservoirs, such as heavy oil and shale oil. During the last two decades, the development of thermal stimulation technologies, such as in-situ combustion, cyclic steam injection and SAGD (steam assisted gravity drainage), have played a major role in the implementation of different concepts of oil production from unconventional oil reserves. Today, about 60% of world oil production attributed to methods of enhanced oil recovery (EOR) comes from thermal stimulation (Chekhonin et al., 2012). As an alternative to traditional thermal methods to reduce oil viscosity and enhance recovery, electric/electromagnetic heating method has been proposed and tested in heavy oil and oil shale reservoirs (Sahni, 2000; Hascakir, 2008), to convert oil shale to producible oil and gas through heating the oil shale in situ by hydraulically fracturing the oil shale and filling the fracture with electrically conductive material to form a heating element (Symington et al., 2006), as shown in **Fig.1**. Already, microwave applications in oil sands bitumen and shale oil production and in petroleum upgrading are gaining considerable interest in recent years. Energy companies and petroleum researchers have been working on a variety of unconventional technologies such as microwave and radio frequency (RF) energies to recover viscous oil from shale (Mutyala et al., 2010), and it is observed that microwaves can heat up the formation much faster than conventional steam heating. In addition, nanoparticles in the form of nanofluids have been investigated for enhanced oil recovery applications and it has shown that due to absorption of electromagnetic waves by the cobalt ferrite nanoparticles, oil viscosity can be reduced, resulting in a significant increase of oil recovery (Yahya et al., 2012).

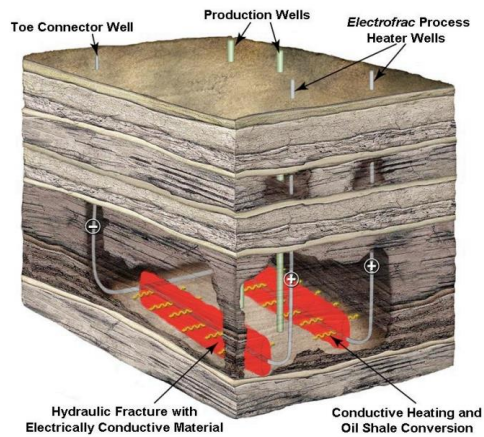

**Fig. 1—Heating hydraulic fracture with electrically conductive material (Hoda et al., 2010)**

Even though numerous studies have investigated how to improve heavy oil/shale oil recovery by increasing formation temperature, limited studies have explored the possibility of enhancing gas recovery with similar thermal stimulation techniques. Salmachi et al. (2012) investigated the feasibility of enhancing gas recovery from coal seam gas reservoirs using geothermal resources. The results show that hot water injection for a period of 2 years into the coal seam with an area of 40 acres successfully increases average reservoir temperature by 30 ℃ and dramatically increases gas recovery by 58% during 12 years of production. Wang et al. (2015a) also concluded that thermal stimulation has the potential to enhance CBM recovery substantially by liberating a significant amount of residual adsorption gas. Wang et al. (2014) investigated the application of thermal stimulation in hydraulically fractured shale gas formations by altering gas adsorption/desorption behavior in multiple transverse hydraulic fractures. Their work shows the efficiency of thermal treatment in shale gas formations largely depends on fracture spacing, operation conditions, gas adsorption and rock properties. Chapiro and Bruining (2014) investigated the possibility of in-situ combustion to improve permeability in shale gas formations. They concluded that if kerogen present in sufficient quantities, methane combustion can generate enough heat to enhance the permeability. Yue et al. (2015) conducted laboratory experiment on shale samples, by measuring shale gas adsorption capacity with different temperatures, their work indicates that large amount of adsorption gas can be expelled by elevating temperature and the sensitivity of gas adsorption capacity to temperature depends on rock mineralogy properties.

The potential application of thermal stimulation in shale gas formations adds extra complexity in reservoir simulation and modeling, because the rising temperature can substantially impact real gas properties, the release of adsorption gas, pressure depletion process and the associated matrix and fracture flow capacity. So it is crucial to have a reservoir simulation model that can deal with these multi-physics problems with enough confidence, however, no reservoir simulation model, that is able to encompass all the coupled physics, is currently available and no literature has attempted to do so.

In this study, we propose a fully coupled thermal-hydraulic-mechanical (THM) model for fractured shale gas reservoirs and evaluate the feasibility of using thermal stimulation in hydraulically fractured shale formations to enhance the ultimate recovery. The results of this study provide us a better understanding of the coupled behavior of fluid flow, rock deformation and heat transfer under complex reservoir conditions, and help us assess the effectiveness and potential applications of thermal stimulation methods in fractured shale gas formations. The structure of this article is as follows. First, a brief description of shale gas reservoir THM modeling will be presented. Then, this fully coupled model is applied to different thermal stimulation scenarios and the effects the thermal stimulation are assessed. Finally, conclusion remarks and discussions are presented. Detailed mathematic formulations for the proposed model are discussed in **Appendix**.

## Thermal-Hydraulic-Mechanical Modeling in Fractured Shale Gas Reservoirs

Be able to model and simulate reservoir depletion and well performance properly in fractured formations is crucial for shale gas field development. Geological modeling and reservoir simulation provide essential information that can help reduce risk, enhance field economics, and ultimately maximize gas reserves by identifying the number of wells required, the optimal completion design and the appropriate enhanced recovery methods. The reliability of reservoir simulation results is primarily dependent on the quality of input parameters and the physical modeling of the reservoir-production systems. And the reservoir simulation model itself should be robust, representative and be able to make the most use of available data that come from various sources. Besides extremely low formation matrix permeability, some other unique features of shale gas reservoir, which can substantially impact the reservoir simulation results, should not be ignored.

In classic fluid flow mechanics where continuum theory holds, fluid velocity is assumed to be zero at the pore wall (Sherman, 1969). This is a valid assumption for conventional reservoirs having pore radii in the range of 1 to 100 micrometer, because fluids flow as a continuous medium. Correspondingly, Darcy's equation, which models pressure-driven viscous flow, works properly for such reservoirs. However, in shale reservoirs, the ultrafine pore structure of these rocks can cause violation of the basic assumptions behind Darcy's law. Depending on a combination of pressure-temperature conditions, pore structure and gas properties, Non-Darcy flow mechanisms such as Knudsen diffusion and Gas-Slippage effects will impact the matrix apparent permeability (Fathi et al. 2012; Michel et al. 2011; Swami et al. 2012; Sakhaee-Pour and Bryant, 2012). In addition, constant decreasing pore pressure during production (transient flow and pseudo-steady-state flow) can lead to reduction of thickness in gas adsorption layer and increase in the effective stress, which in turn, can impact the formation matrix microstructure and effective pore radius (Wang and Marongiu-Porcu, 2015). So the overall matrix apparent permeability in shale gas reservoir is dynamic and pressure dependent, as shown in **Fig.2**. In the context of hydraulic fractured shale gas formations, the local matrix permeability in the Stimulated Reservoir Volume (SRV) is space and time dependent during production.

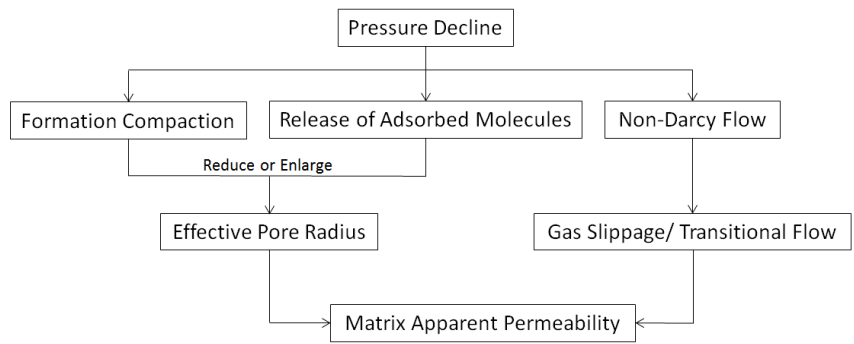

**Fig. 2—Mechanisms that alter shale matrix apparent permeability during production** (Wang and Marongiu-Porcu, 2015)

Besides the matrix permeability, the surface area of natural fracture networks that connected to the main hydraulic fracture and its ability to sustain conductivity are also critical for predicting long-term production in shale gas formations (Ghassemi and Suarez-Rivera, 2012). It is recommended that Brinell Hardness Test (BHN) and Unpropped Fracture Conductivity Test (UFCT) should be done in shales in order to determine fracture treatment types and estimate the relationship between fracture conductivity and confining stress (Ramurthy et al. 2011). So the effects of pressure dependent matrix permeability and fracture conductivity should be included in the simulation model for any well performance and production prediction.

The inclusion of temperature effects add more complexity in reservoir modeling and simulation, because the changes in formation temperature not only alter gas adsorption capacity, vary real gas properties, induce thermal stress, but also impact matrix permeability and fracture conductivity in a fully coupled manner, which in turn, affect in-situ flow capacity and ultimate recovery. **Fig.3** shows laboratory measurement and theoretical model prediction of gas adsorption capacity at different temperatures from a shale sample (Yue et al., 2015). By examining the adsorption curves, we can deduce that the reservoir pressure must be sufficiently low to liberate the adsorbed gas and the ultimately recoverable amount of gas is largely a function of the adsorbed gas that can be released (desorbed). Because most adsorbed gas can only be released at low reservoir pressure, due to the extremely low permeability in shale matrix, even with hydraulic fracturing, it would take considerable production time for the average pressure within the drainage area to drop to a level where most of the adsorbed gas can be liberated, and the production rate may have already reached the economical shut-in limits by then. However, if the reservoir temperature can be elevated, a significant amount of adsorption gas can be released, even at high pressure (golden line in Fig.3). Thus, thermal stimulation techniques can be utilized as a potential method to enhance ultimate recovery from shale gas reservoir by altering shale gas desorption behavior.

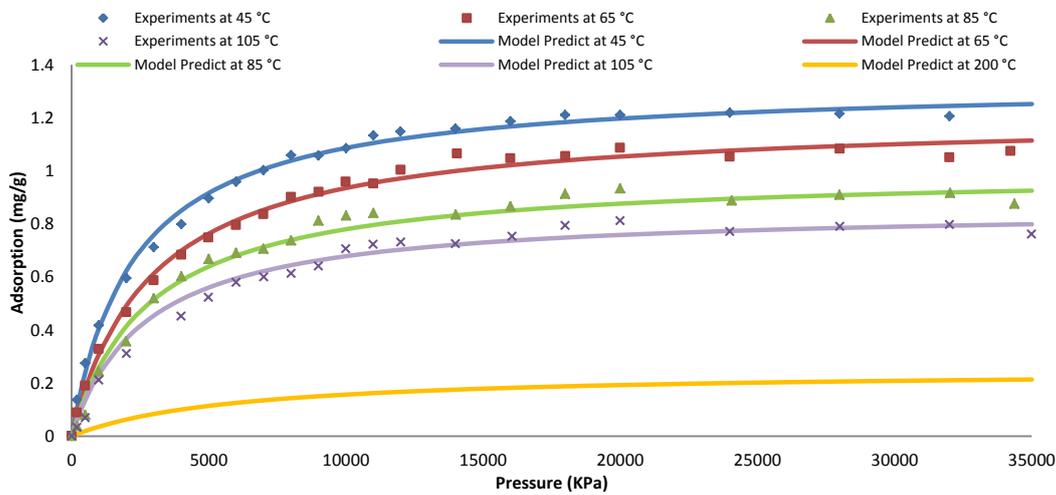

**Fig. 3—Laboratory measurement and Bi-Langmuir model prediction of gas adsorption capacity at different temperature for a shale sample (Yue et al., 2015)**

With the elevated formation temperature, thermal stimulation induced rock mechanical behavior can alter rock properties and hence impact the fluid flow. Numerous studies (Burghignoli et al., 2000; Cekerevac et al., 2004, Laloui et al., 2003; Wong et al., 2006; Xu et al., 2011; Yuan et al., 2013) have investigated the behavior of different rocks during the process of thermal stimulation, indicating that the rocks can be weakened or strengthened depending on the factors such as initial porosity, applied confining pressure, heating rate and rock composition. However, the relevance of these results to field enhanced recovery project using thermal stimulation remains poorly understood. So in this article, we do not consider dynamic rock damage mechanisms during thermal stimulation, but effects of induced thermal stress on associated, coupled physics are well captured. A comprehensive discussion on THM modeling and mathematical formulations are presented in **Appendix**.

## Model Description

In this section, the application of the proposed THM model is presented and discussed. Even though it is possible to simulate an entire section of a horizontal well with multiple transverse fractures, it is more efficient to simulate a unit pattern and apply symmetric boundary conditions along boundary of SRV that contains one main hydraulic fracture and natural fractures, as shown in **Fig. 4** (the red line depicts horizontal wellbore, the blue line, and the distributed black line segments represent hydraulic fracture and natural fractures respectively).

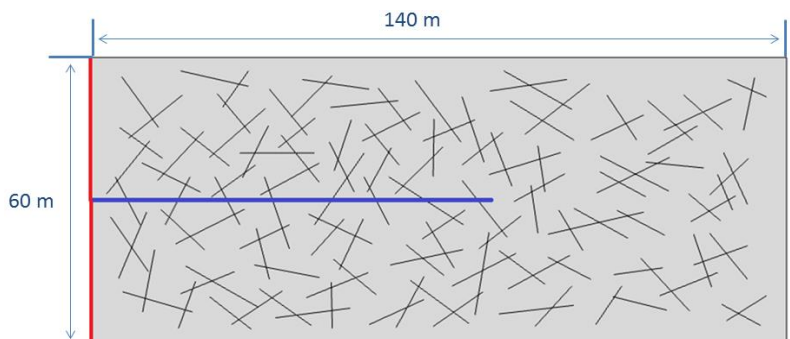

**Fig.4—A plane view of SRV containing hydraulic fracture and natural fractures**

In practice, the primary hydraulic fracture and secondary fracture networks within each SRV unit, can be created statistically by using seismic, well and core data (Ahmed et al. 2013; Cornette et al. 2012 ) and by matching hydraulic fracture propagation models. Hydraulic fracture models that commonly used in industry relies on linear elastic fracture mechanics (LEFM) (Johri and Zoback 2013; Weng 2014), but it is only valid for brittle rocks (Wang et al. 2016), recent studies demonstrate that using energy based cohesive zone model can resolve the issue of singularity at the fracture tip and these cohesive zone based hydraulic fracture models not only can be used in both brittle and ductile formations, but also can capture complex fracture evolution, such as natural fracture interactions (Guo et al. 2015), fracture reorientation from different perforation angles (Wang 2015), producing well interference (Wang 2016a), the effects of fracturing spacing and sequencing on fracture interaction/coalescence from single and multiple horizontal wells (Wang 2016b).

A limitation of the widely used commercial reservoir simulators is the fracture usually modeled as a narrow space with high porosity/permeability. A grid block that contains fracture segment is given higher porosity/permeability. The smallest grid block is often much thicker than typical fracture width. This requires the fracture property to be homogenized within the containing grid blocks. The orthogonal grid based mesh, used by many reservoir simulators, also imposes some limitation on fracture shape and geometries. In order to avoid extreme mesh refinement inside and around the fractures and guaranteeing solution convergence, discrete fracture networks (DFN) and unstructured mesh are implemented in our proposed model to delineate the fluid flow behavior along the fractures and discretize reservoir domain. **Fig. 5** shows the unstructured mesh generated in the simulation domain that discretizes both reservoir and discrete fracture networks. The maximum and minimum element size is 0.5 m and 0.1 m, respectively and there are 96574 elements in total.

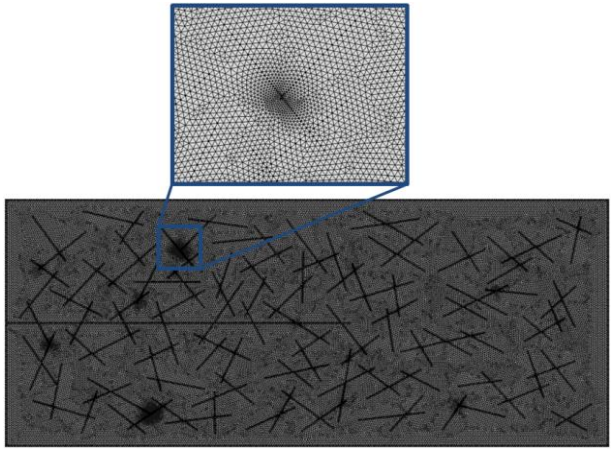

**Fig.5— Unstructured mesh and DFN distribution within simulation domain**

Combine hydraulic fracture modeling and microseismic data, the spatial distribution of fracture networks can be estimated by various methods (Johri and Zoback, 2013; Yu et al., 2014; Aimene and Ouenes, 2015). And the properties of the propped fracture and un-propped fracture can be determined through conductivity test under various pressures and confining stresses (Ghassemi and Suarez-Rivera, 2012), so that the parameters define transport capacity of fractures can be determined (e.g., fitting Eq.A-20 with laboratory data). With all these information, the special and intrinsic properties of DFNs can be estimated. How to map fracture networks across different fracturing stages based on microseismic data/hydraulic fracture modeling and fully characterize the entire horizontal completion is beyond the scope of this article. The distribution of natural fractures is randomly assigned within SRV in this study as pictured in Fig. 5.

**Table 1** shows all the input parameters used for the synthetic case analysis, which includes reservoir conditions, drainage geometry, fracture conductivity, real gas parameters and rock thermal properties for a typical shale formation. The diameter of adsorption gas molecules is assumed to be 0.414 nm, which is the diameter of a methane molecule. The combination of input parameters is designed to reflect a scenario where 60% of initial gas in-place comes from adsorbed gas, this can be the case in many organic-rich shale formations (Darishchev et al., 2013).

| Input Parameters | Value |
|---|---|
| Initial reservoir pressure, $p_0$ | 38[MPa] |
| Wellbore pressure, $p_{wf}$ | 5 [MPa] |
| Initial intrinsic matrix permeability, $k_{\infty 0}$ | 100[nD] |
| Initial formation porosity, $\emptyset_{m0}$ | 0.01 |
| Initial matrix pore radius (not include adsorption layer), $r_0$ | 3[nm] |
| Porosity compaction parameter, $C_\emptyset$ | 0.035 |
| Initial hydraulic fracture permeability, $k_{f0}$ | 10000[mD] |
| Hydraulic fracture porosity, $\emptyset_f$ | 0.5 |
| Hydraulic fracture width, $d_f$ | 0.01[m] |
| Fracture compaction parameter for hydraulic fracture, $B_f$ | 0.005[1/ MPa] |
| Initial natural fracture permeability, $k_{nf0}$ | 400[mD] |
| Natural fracture porosity, $\emptyset_{nf}$ | 0.01 |
| Natural fracture width, $d_{nf}$ | 0.0001[m] |
| Fracture compaction parameter for hydraulic fracture, $B_{nf}$ | 0.05[1/ MPa] |
| Initial reservoir temperature, $T_0$ | 366[K] |
| Thermal stimulation temperature, $T_s$ | 478[K] |
| Formation heat capacity, $C_m$ | 1,000[J/K/kg] |
| Formation heat conductivity, $\lambda_m$ | 4[W/m/K] |
| Density of formation rock, $\rho_m$ | 2600 [kg $/m^3$] |
| Langmuir volume constant, $V_L$ | 0.0113[$m^3$/kg] |
| Langmuir pressure constant, $P_L$ | 10 [MPa] |
| Diameter of adsorption gas molecules, $d_m$ | 0.414[nm] |
| Average gas molecular weight, $M$ | 16.04[g/mol] |
| Critical temperature of mix Gas, $T_c$ | 191 [K] |
| Critical pressure of mix gas, $P_c$ | 4.64[MPa] |
| Hydraulic fracture spacing, $Y_e$ | 60[m] |
| Hydraulic fracture half length, $x_f$ | 80 [m] |
| SRV drainage length parallel to the hydraulic fracture, $X_e$ | 140[m] |
| Reservoir thickness, H | 50[m] |
| Total number of SRV unit, $n_{SRV}$ | 20 |
| Horizontal minimum stress, $S_h$ | 40[MPa] |
| Horizontal maximum stress, $S_H$ | 45[MPa] |
| Overburden stress, $S_v$ | 50[MPa] |
| Biot's constant, $\alpha$ | 1 |
| Thermal expansion coefficient, $\beta$ | 0.00003[1/K] |
| Young's modulus, $E$ | 25[GPa] |
| Poisson's ratio, $\nu$ | 0.25 |

Table 1. Input parameters for simulation cases

Bi-Langmuir model (Lu et al. 1995) is used to depict pressure-temperature dependent gas adsorption capacity. By regression on common gas adsorption data (i.e., Langmuir volume constant $V_L$ and Langmuir pressure constant $P_L$), gas adsorption capacity at different temperatures can be extrapolated (Wang et al. 2014). **Fig.6** shows the prediction of gas adsorption capacity at stimulation temperature (478 K) by fitting original Langmuir curve (determined by $V_L$ and $P_L$) at initial reservoir temperature (366 K) using Bi-Langmuir model.

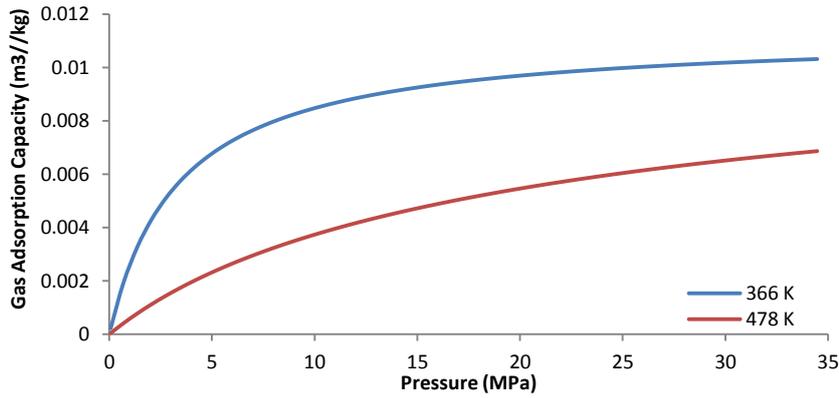

Fig.6—Shale gas adsorption capacity with different pressure and temperature

**Base Case Simulation**

In this section, we present our simulation results for the base case, in which the formation is under isothermal condition throughout the lifetime of reservoir depletion. In the following section, the results from base case simulation will be compared with the results when thermal stimulation is applied. **Fig. 7** shows the pressure distribution in the simulated SRV unit after 1 and 20 years of production. It can be observed that the pressure disturbance has reached the boundary after 1 year's production, owing to the existence of conductive fracture networks, and pseudo-steady-state flow has already ensued from the initial transient flow, to dominate the rest of the production history. After 20 years of production, most of the area that penetrated by the main hydraulic fracture has been completely depleted due to the well-connected fracture system and effective linear flow from matrix to fracture, but the area in front the main hydraulic fracture tip has not been depleted effectively.

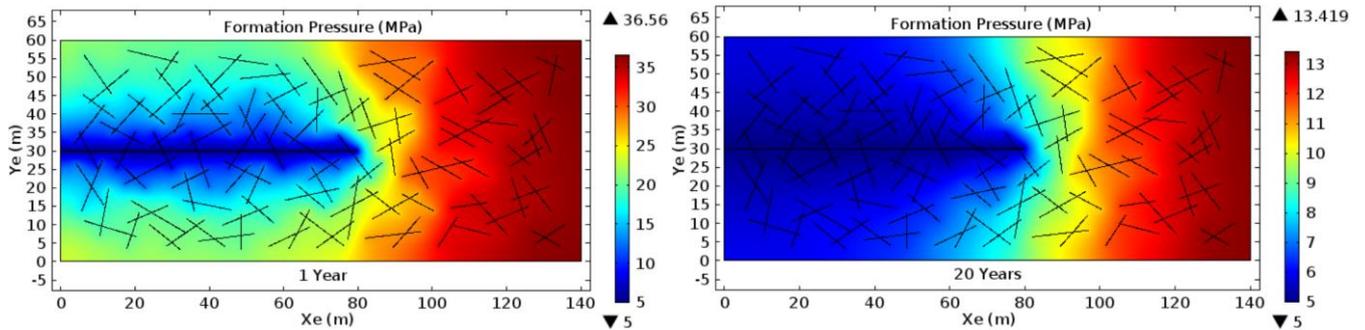

Fig.7—Pressure distribution in the SRV after 1 and 20 years of production

**Fig. 8** and **Fig.9** depicts the gas density and gas viscosity distribution in the simulated SRV unit after 1 and 20 years of production. It is within our expectation that under isothermal reservoir conditions, both gas density and gas viscosity are correlated to reservoir in-situ local pressure. The expansion of gas volume with decreasing gas density during depletion is the main driven mechanism of production, and the gas viscosity decreases as pore pressure declines. We can also observe that the gas viscosity is reduced locally to around 50% of its initial value in low-pressure zone, so if constant gas viscosity is used in reservoir simulation or production prediction models, the inaccurate results may lead to erroneous decisions.

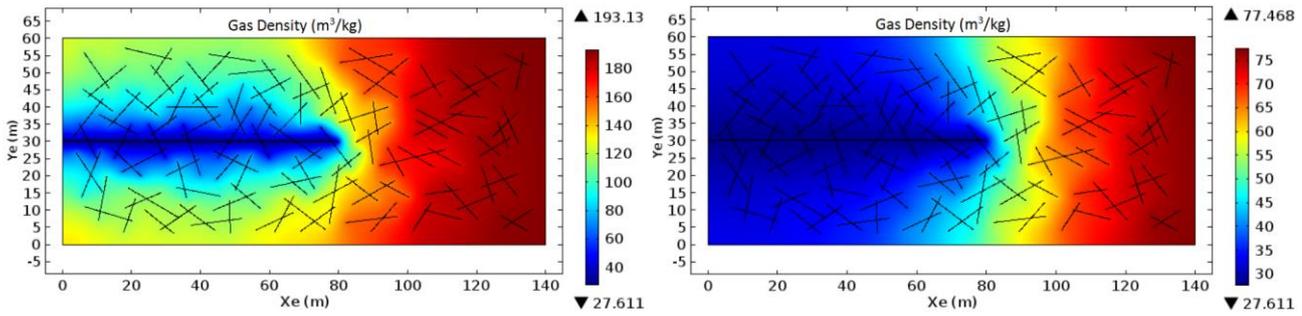

**Fig.8—Gas density distribution in the SRV after 1 and 20 years of production**

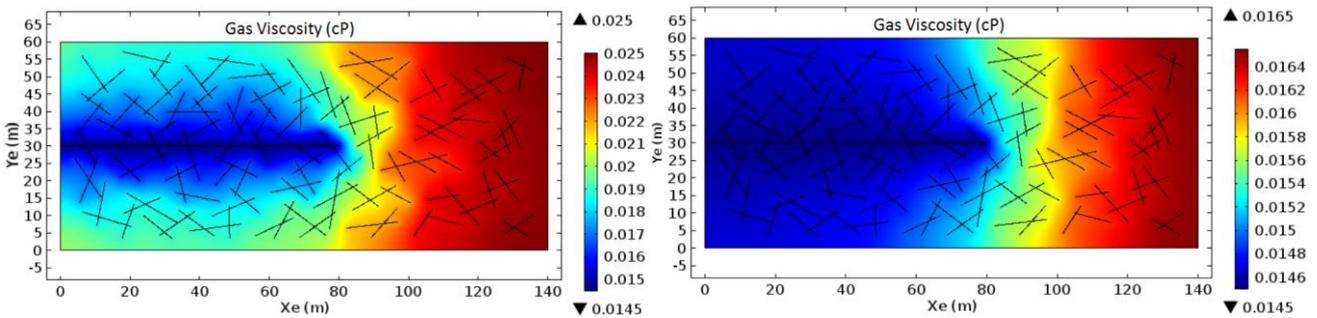

**Fig.9—Gas viscosity distribution in the SRV after 1 and 20 years of production**

**Fig.10** shows the Knudsen Number distribution in the simulated SRV unit after 1 and 20 years of production, respectively. Relate to Fig.7, it can be observed that the value of Knudsen Number is highest in the low-pressure zone and lowest in the high-pressure zone. According to the classification in **Table A1**, the values of Knudsen Number fall within the transition regime in the entire SRV, in which the validity of Darcy flow mechanism certainly breaks down.

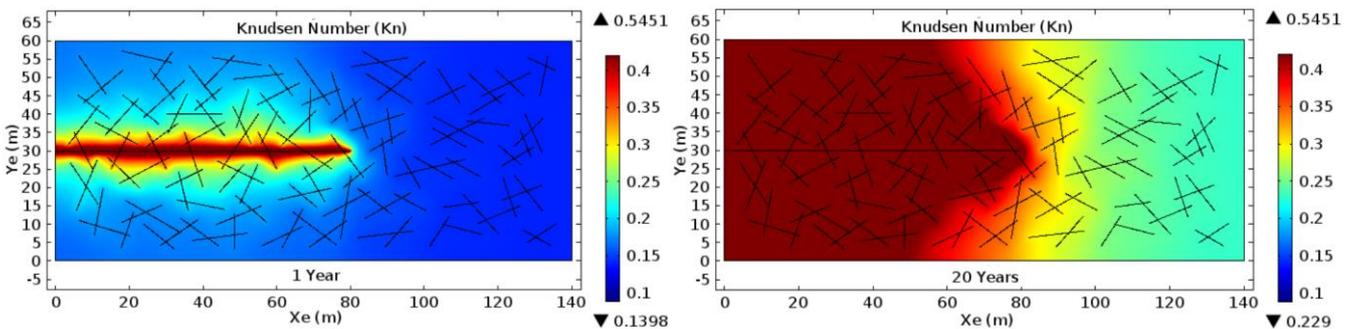

**Fig.10—Knudsen Number in the SRV after 1 and 20 years of production**

The corresponding matrix apparent permeability distribution is shown in **Fig. 11**. It can be observed that the value of matrix apparent permeability is higher within the depleted zone. And as the pressure sink propagating away from the near-hydraulic-fracture-region to the entire SRV, the matrix apparent permeability increases with declining pressure, due to the effects of Non-Darcy flow/Gas-Slippage and the release of adsorption layer in the nano-pore space during production.

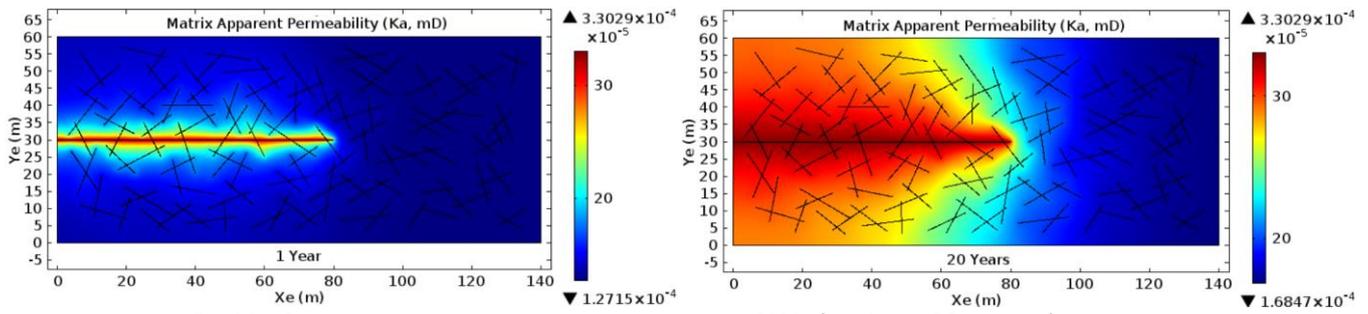

Fig.11—Shale matrix apparent permeability in the SRV after 1 and 20 years of production

**Fig.12** shows how much adsorption gas has been produced within the SRV unit, which can be a good indicator to estimate the ultimate recovery. We can notice that after 20 years of production, nearly 58% of adsorbed gas has been produced in the depleted zone, while only close to 30% of adsorbed gas has been produced in the area in front of the main hydraulic fracture tip. From the simulation results, we can also conclude that the maximum percentage adsorption gas that can be recovered through pressure depletion is around 58%, under our simulated production conditions and gas adsorption/desorption properties that provided in Table 1.

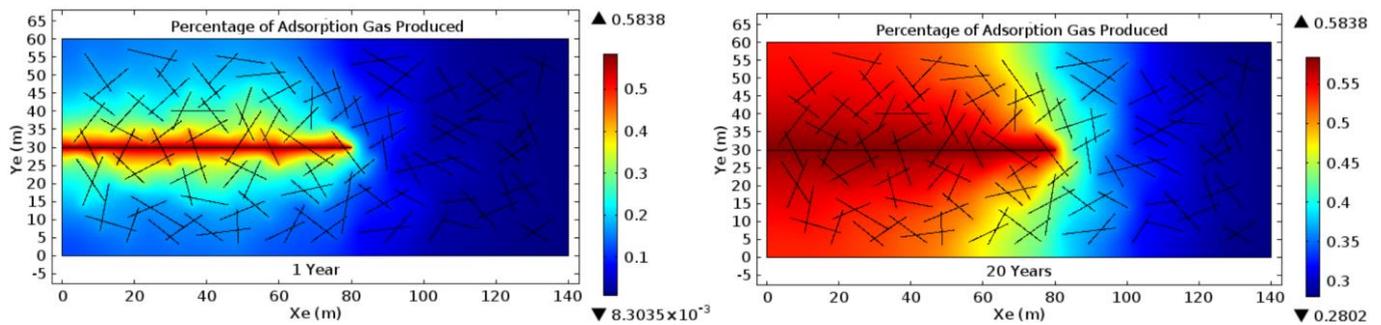

Fig.12—Percentage of adsorption gas produced in the SRV after 1 and 20 years of production

In order to demonstrate the importance of incorporating the evolution of both fracture conductivity and matrix permeability during production, two additional scenarios are constructed with same input parameters as the base case, but assuming constant matrix permeability and constant fracture conductivity, respectively. Because fracture conductivity is the product of fracture permeability and fracture width, so by assuming the fracture width remains the same, shift fracture permeability as a function of local stress (Eq. A.20) has the same effect as shift fracture conductivity on simulation results. In addition, the fracture storage effects have negligible impact on gas production, because the driving forces of primary shale gas recovery come from the gas volume expansion and gas desorption with declining pressure. **Fig.13** shows the predicted cumulative production for three different scenarios. It can be clearly seen that overlook the pressure dependent fracture permeability/conductivity can lead to overestimation of production, while ignoring the pressure dependent matrix permeability can result in underestimation of production. In fact, the relative importance of matrix apparent permeability evolution no only depends on reservoir micro-structure, pressure-temperature conditions, but also influenced by the density, connectivity and conductivity of natural fractures. If the overall flow capacity inside the SRV is dominated by abundant fracture networks, then the general role of matrix permeability itself diminishes (Wang 2016c). Being able to predict production correctly is crucial to optimize field development and reservoir management, so a comprehensive reservoir characterization is required to be able to delineate reservoir performance with enough certainty, because the flawed physical model can render unreliable prediction and poor history match even with available production data.

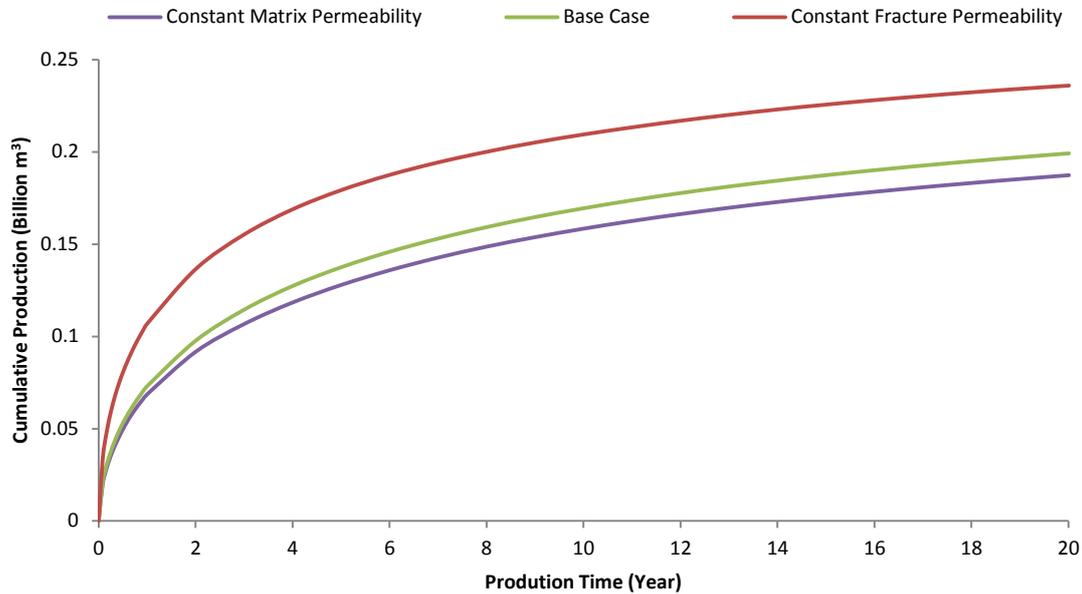

Fig.13—Prediction of cumulative production with different simulation models

**Case 1: Thermal Stimulation through Hydraulic Fracture**

In this section, we simulated the case of directly heating the propped hydraulic fractures, assess the proposed method of using electromagnetically sensitive proppants, and investigate how the increased formation temperature can impact reservoir depletion, cumulative production, and ultimate recovery. In this case, the thermal stimulation temperature is set to be 475K along the main hydraulic fracture and remains constant throughout the production life; all the other input parameters maintain the same as the base case. It can be expected that with elevated formation temperature, more adsorption gas can be released at the same pressure, as illustrated in Fig.6: the extra gas can be produced is equivalent to the difference of shale rock adsorption capacity between two temperature states at a specific local pressure, if do not consider the changes in formation flow capacity. In addition, all the pressure-temperature dependent variable will be affected by the rising temperature, so does the fully coupled THM interactions.

**Fig.14** shows the temperature distribution within the simulated SRV after 1 and 20 years of production. It can be observed that the front edge of stimulation temperature has not reached the boundary yet after 1 year's production and the temperature distribution profile reflects a heat transfer process that dominated by the conduction of rock matrix. The temperature propagation speed depends on formation thermal properties and thermal stimulation temperature. Less time will be needed to heat up the formation temperature to the desired value if the formation exhibits higher thermal conductivity or higher stimulation temperature is imposed. After 20 years of production under current simulation conditions, most of the SRV that penetrated by the hydraulic fracture has reached the desired formation temperature. Compare with the pressure depletion profile in Fig.7, we can infer that the efficiency of pressure depletion and heat transfer is substantially impacted by the length of hydraulic fracture that penetrated into the drainage volume, even with the existence of conductive natural fractures. So unlike conventional reservoirs, it is the length and spacing of hydraulic fracture that determines the actual drainage area in shale gas formations.

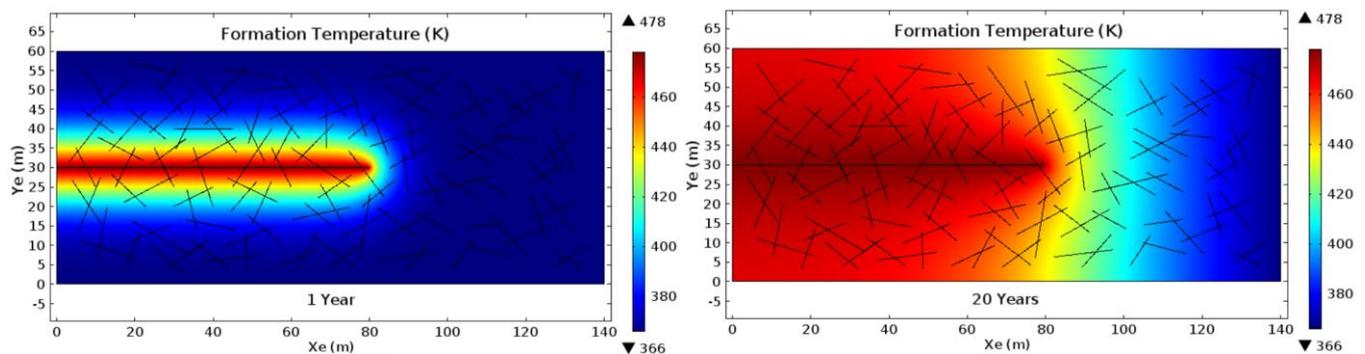

Fig.14—Temperature distribution in the SRV after 1 and 20 years of production

**Fig. 15** shows the distribution of pressure, Knudsen Number, matrix apparent permeability and percentage of adsorption gas produced in the simulated SRV after 20 years of production, with the application thermal stimulation. Compare with previous results from the base case, the impact of elevated formation temperature on gas transport and desorption mechanisms can be clearly observed: The average pressure in the SRV (Fig. 15-a) is higher than that in the base case (Fig.7) at the end of 20 years production when thermal stimulation is implemented, especially within the regime in front of the hydraulic fracture tip. This can be explained by the fact that more adsorption gas is released due to increasing formation temperature, which in turn compensates parts of the pressure depletion and leads to a lower rate of reservoir pressure decline. Compare Fig. 15-b with Fig. 10, we can note that the Knudsen Number increases from 0.5451 to 0.8151 in the SRV regimes where the hydraulic fracture penetrates. This is because the temperature and pressure dictate the alteration of Knudsen Number during production and the elevated temperature contributes to higher Knudsen Number (as reflected by Eq.A.21). However, within the regime in front of the hydraulic fracture tip, the lowest Knudsen Number even drops from 0.229 (base case) to 0.1909 (with thermal stimulation). This contrast effects of thermal stimulation in different regimes is due to the fact that, unlike the regime penetrated by the hydraulic fracture, the regime far ahead of the hydraulic fracture tip has not been impacted temperature propagation yet (Fig. 14), and the pressure in this regime is higher because pressure depletion has been compensated by the release of extra adsorption gas in the regime where temperature is elevated. This opposite effects of thermal stimulation in these two divided regions also applies to the matrix apparent permeability and the percentage of adsorption gas produced. Higher Knudsen Number in the hydraulic fracture penetrated SRV indicates more severe the Non-Darcy flow/Gas-Slippage behavior. In addition, elevated temperature also lead to higher intrinsic permeability resulted from thinner adsorption layer, that's why the maximum matrix apparent permeability increases from 330-nd (base case) to 461-nd (with thermal stimulation) in the simulated SRV, as we can discern from Fig. 15-c and Fig. 11. More interestingly and importantly, when comparing Fig. 15-d with Fig. 12, it can be observed that the maximum recovery of the adsorption gas in the SRV increased from 58.38% to 81.33% with the help of temperature elevation.

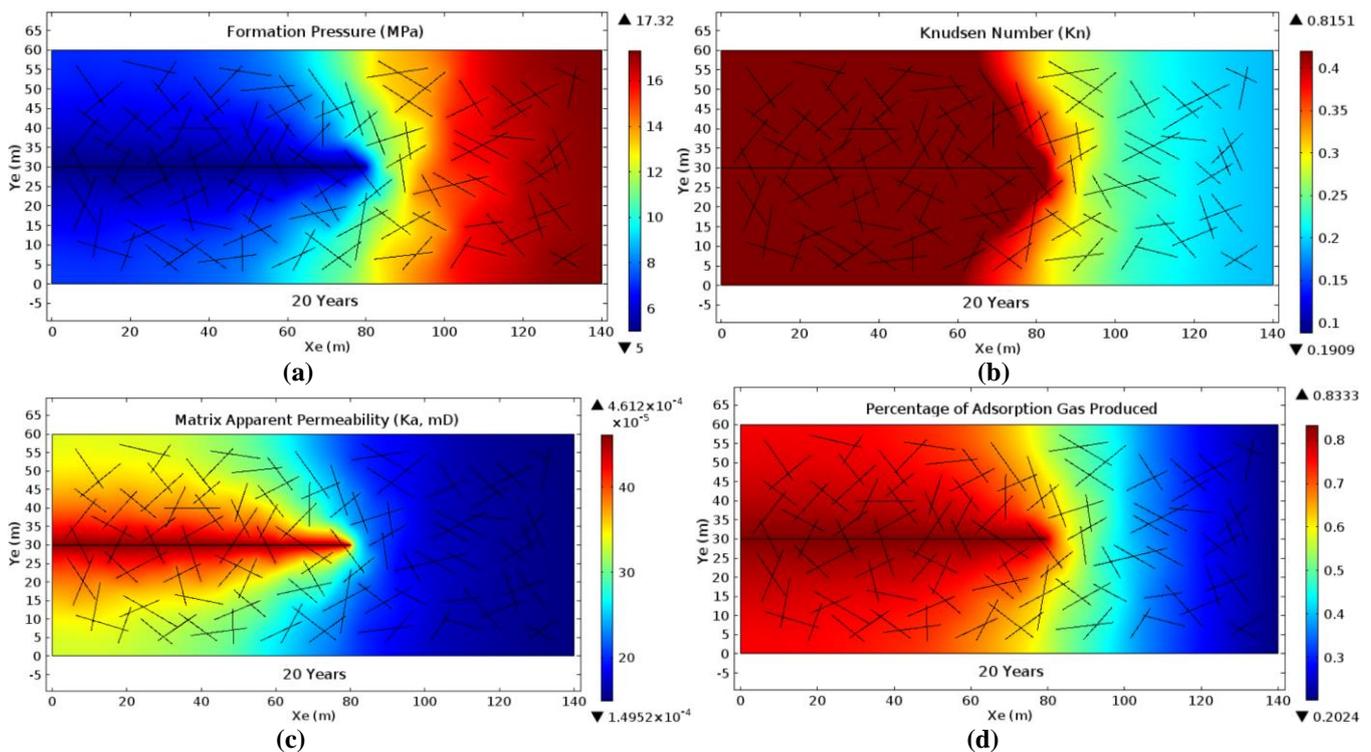

**Fig.15—The distribution of pressure (a), Knudsen Number (b), matrix apparent permeability (c) and percentage of adsorption gas produced (d) in the SRV after 20 years of production with thermal stimulation**

**Fig.16** shows the predicted cumulative production for the base case and the thermal stimulated case. The cumulative production after 20 years increases 40% when thermal stimulation is applied along the hydraulic fracture, with 122K difference between the thermal stimulation temperature and initial reservoir temperature. With elevated formation temperature, more adsorption gas can be released ultimately under current well production conditions. In addition, higher temperature leads to higher matrix apparent permeability due to stronger Non-Darcy flow/Gas Slippage behavior and slower pressure decline trend, which in turn, can relax fracture permeability reduction. On top of that, more gas can be expelled from the reservoir because of gas volume expansion at a higher temperature. All these factors lead to higher production and enhanced ultimate recovery.

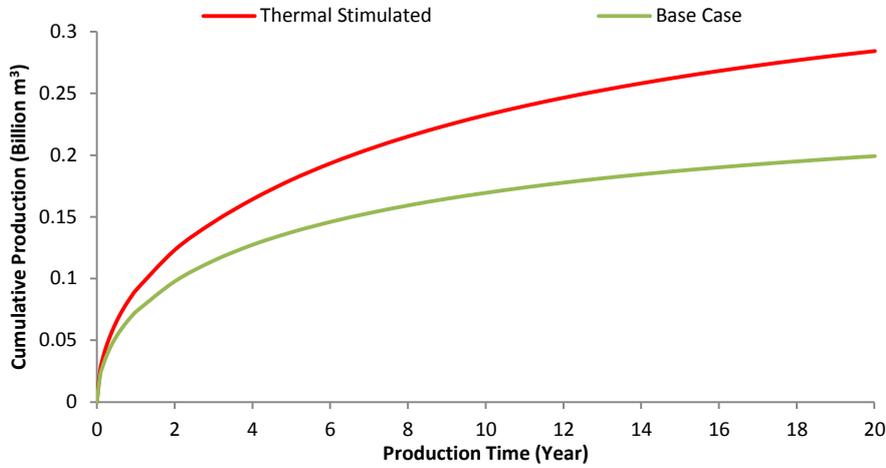

**Fig.16—Prediction of cumulative production with and without thermal stimulation, $k_{\infty 0}$=100 nd**

Next, we investigate how shale matrix permeability can influence thermal stimulation efficiency. Low formation permeability adversely impact the pressure diffusion process and in such cases, the transient flow in SRV may dominate the entire production life and the average pressure in the SRV will not be able to decline to a level that allows sufficient desorption before production rates reach economical limits. Here we reduce the initial intrinsic matrix permeability from 100 nd to 10 nd and maintain other input parameters as the same as provided in Table 1. **Fig. 17** and **Fig.18** depict the formation pressure and mean effective stress in the SRV after 1 and 12 months of production, respectively. We can observe that after 1 month of production, only the regimes that around natural fractures that well-connected to the main hydraulic fracture are depleted, this is due to the low permeability of the formation and the fracture networks serve as the main path for gas flow. Also, the asymmetric nature of fracture network distribution leads to the disproportion of depletion with the SRV on both sides of the hydraulic fracture after 1 year of production. With constant far field stress, the local mean effective stress evolves with pressure depletion.

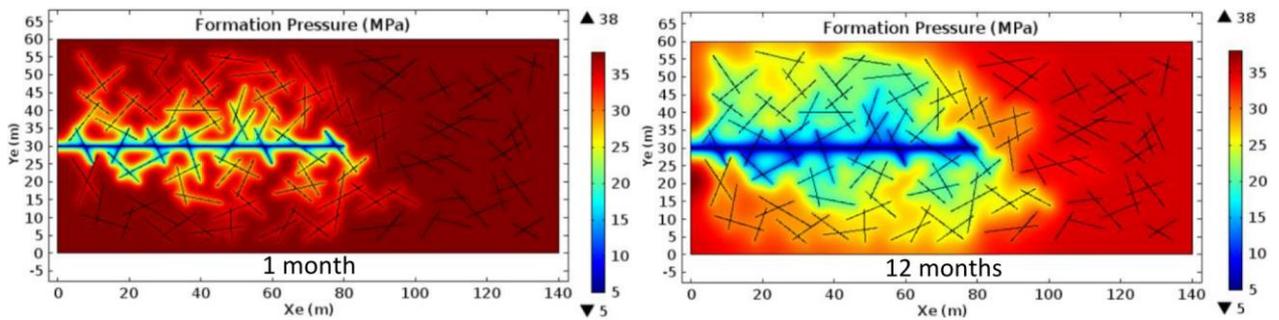

**Fig.17— Pressure distribution in the SRV after 1 and 12 months of production**

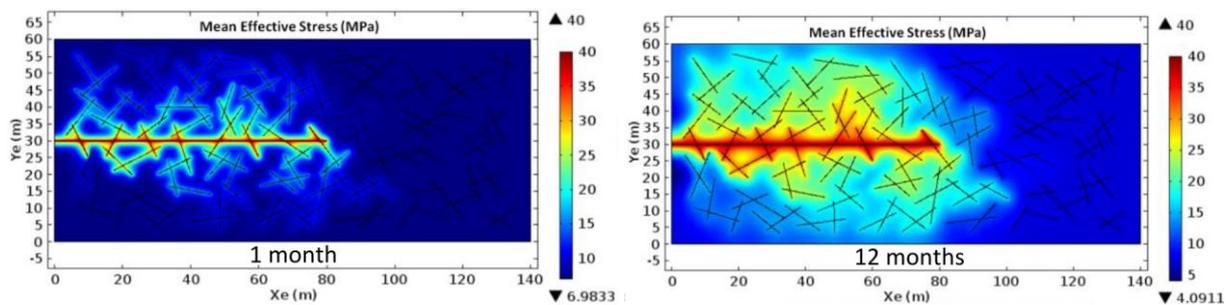

**Fig.18— Mean effective stress distribution in the SRV after 1 and 12 months of production**

**Fig.19** shows the predicted cumulative production for the base case and the thermal stimulated case, with the same initial matrix intrinsic permeability of 10 nd. The cumulative production after 20 years increases 50% when thermal stimulation is applied along the hydraulic fracture. Compared to Fig.16, where around 40% of extra gas can be produced with thermal aid, it seems that thermal stimulation is more efficient when formation permeability is lower. This can be explained by the fact that the lower the formation permeability, the less effective of pressure depletion, so temperature elevation has more impact on gas production.

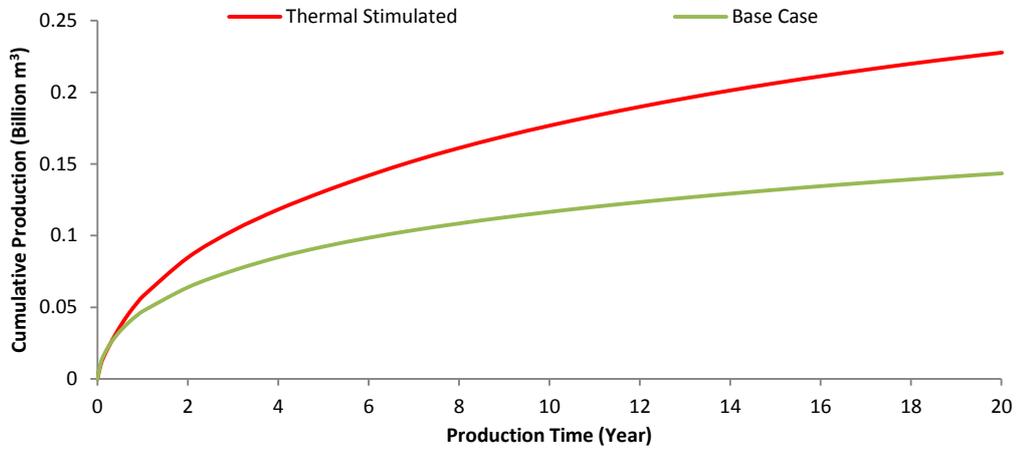

**Fig.19—Prediction of cumulative production with and without thermal stimulation, $k_{\infty 0}$=10 nd**

Even though, thermal stimulation has the potential to enhance gas recovery significantly, but the time requires elevating formation temperature to the target level is substantial, because of low efficiency of heat transfer by pure conduction (as shown in Fig.14), so the effect of thermal stimulation on ultimate recovery can be only prominent in late times. Considering a typical shale gas production decline trend, shown in **Fig.20**, where shale gas wells lose 80% of their initial production rate in the first 2 years, and the economically preferred completion designs may be more driven by the net present value derived in the first a few years of production rather than the ultimate recovery of the well. So for practical applications of thermal stimulation techniques in shale gas reservoirs, the heat transfer efficiency has to be improved.

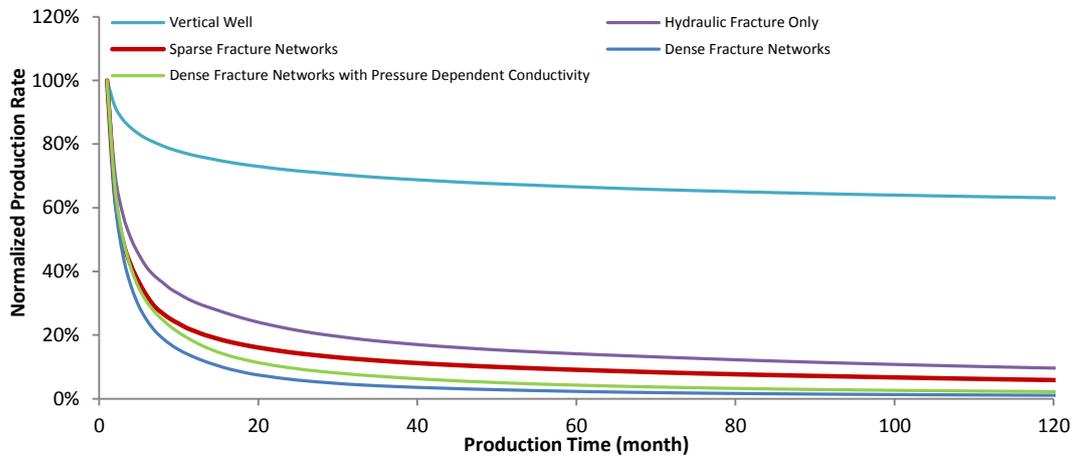

**Fig.20—Typical shale gas production decline trend under different scenarios (Wang, 2016c)**

**Case 2: Thermal Stimulation through Fracture Networks**

The process of proppant transport and settling in complex fracture networks is a complicated phenomenon, where the final distribution of proppant depends on a series of factors, such as fracture geometry, pump rate, proppant concentration, proppant size and injection fluid rheology. In reality, only a fraction of the fracture networks in the SRV is propped and the effective propped volume (EPV) may play a dominant role in determining long-term well performance. **Fig.21** shows the reservoir drainage prediction in SRV and EPV after 20 years of production (the fracture network is shown in black, the virgin pressure is red, SRV in blue rectangular and EPV in green rectangular). It can be noted that most drainage happens inside EPV. Despite ongoing efforts to enhance our understanding of proppant transport in complex fracture geometry (Raymond et al., 2015; Roy et al., 2015; Sahai et al., 2014), how to correctly quantifying proppant distribution within the fracture network on a field scale is still poorly understood and further investigations are required.

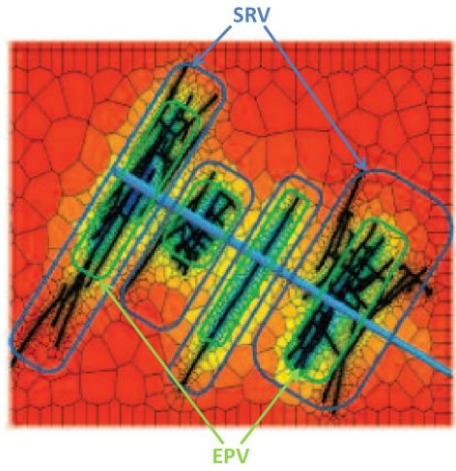

**Fig.21—Simulation of reservoir drainage after 20 years of production (Maxwell, 2013)**

In this section, we investigate a scenario where the heat source particles can be pushed into fracture networks, so more formation area can be heated up simultaneously. To this end, we assume that all the natural fractures that well connected to the hydraulic fracture can be filled with heat source particles, which is the upper-limit case on how large the fracture surface area can be thermally stimulated at the same time. The intention of this section is not to mimic proppant distribution in a realistic field case, but to assess the impact of heat transfer efficiency on gas production and the effectiveness of thermal stimulation.

**Fig.22** shows the temperature distributions in the SRV under such conditions. Compared to Fig. 14, where only the hydraulic fracture acts as heat source, the heat transfer efficiency increases dramatically. When most of the natural fracture networks can be thermally stimulated collectively, the whole SRV regime that penetrated by the main hydraulic fracture can achieve the desired temperature level around 1 year.

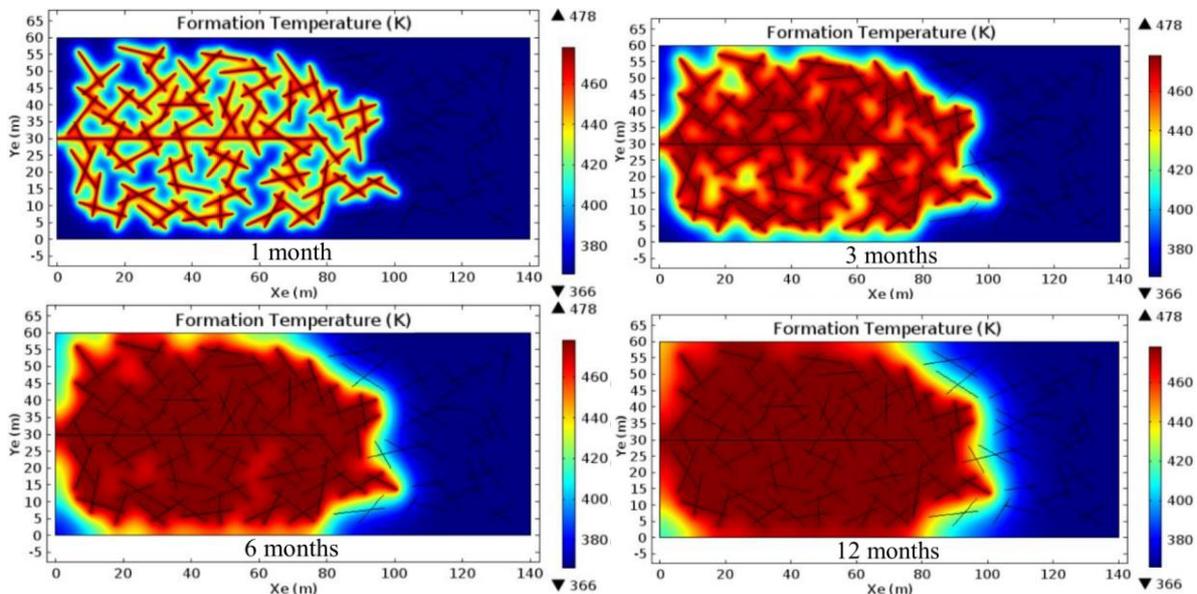

**Fig.22— Temperature distribution in SRV when well-connected fracture networks are thermal stimulated simultaneously**

With such high efficiency of heat transfer, the overall thermal stimulation effects on gas production can be altered significantly, as reflected in **Fig. 23**. When compared to Fig.16, where only the main hydraulic fracture is thermal stimulated, the effect of thermal stimulation becomes prominent (cumulative production almost doubled) even within the first 5 years of production. However, as time goes on, the effects of thermal stimulation gradually tapers off as most the formation have already reached the target temperature within the first a few years. This result demonstrate that if the efficiency of heat transfer process can be improved to desorb large amount of adsorption gas and enhance the overall flow capacity during the initial stage of gas producing well, then the ultimate recovery during the economic lifetime of a gas producing well (typically, first 5 years of production matters most) can be substantially enhanced.

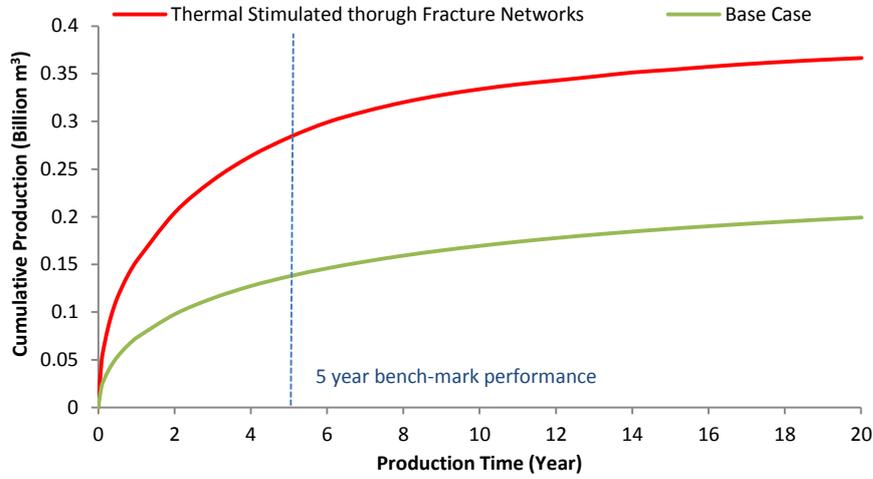

**Fig.23—Prediction of cumulative production with and without thermal stimulation, $k_{\infty 0}$=100 nd**

### Energy Balance Calculations

Different thermal stimulation methods have different capital cost, that related to the maturity of associated technology, the supply of raw material, the efficiency of management and the environmental impact, etc. And the price of natural gas, like other commodities, has the nature of volatile and cyclicity. So, to present a comprehensive evaluation from the economic aspects of thermal stimulation project is out of the scope of this article. Nevertheless, we can still assess the feasibility of thermal stimulation in shale formations by investigating the energy needed for thermal stimulation and the corresponding extra energy output.

The heat content of natural gas, or the amount of energy released when a volume of gas is burned, varies according to gas compositions. The primary constituent of natural gas is methane, which has a heat content of 1,010 British thermal units per cubic foot (Btu/cf), equivalent to $3.763 \times 10^7$ J/m³, at standard temperature and pressure.

**Fig.24** shows the heat content of natural gas consumed in various locations in the United States, and the average heat content of natural gas in the United States was around 1,030 Btu/cf ($3.838 \times 10^7$ J/m³). In the following calculations, we use the heat content of pure methane as a reference value to determine the volume of natural gas needed to generate an equivalent amount of input energy for thermal stimulation.

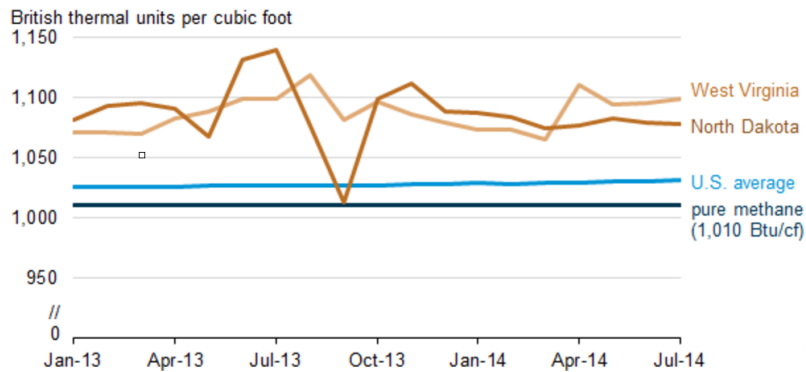

**Fig.24—Heat content of natural gas consumed in various locations in the United States (U.S Energy Information Administration, 2014)**

The energy supplied during thermal stimulation can be estimated using the following equation:

$$\text{Energy Input} = \iiint_V \rho_m C_m (T - T_0) \qquad (1)$$

where $\rho_m$ and $C_m$ are formation rock density and heat capacity, respectively. $T_0$ is the reservoir initial temperature and T the is rock current local temperature. Integrate $\rho_m C_m (T - T_0)$ over the entire simulated volume, the total energy input can be determined.

Based on simulation results from previous sections and energy input and output calculations, the net extra natural gas produced via thermal stimulation can be estimated, as reported in **Table 2**. From the results of case 1, where only the surface of hydraulic fracture is heated, we can observe that the net recovery of extra gas increases as time goes on, regardless of matrix permeability. In absolute terms, more net extra gas can be produced when permeability is higher, but in percentage terms (compared to the scenario where thermal stimulation is not applied), thermal stimulation seems more efficient when permeability is lower. One can also notice that the calculated energy inputs in case 1 are the same, even for different matrix permeability, this is because only conductive heat transfer is modeled, due to the negligible convective heat transfer effects by gas ( refer to Eq.A.11-A.14). The consequences can be quite different if all the well-connected fracture networks can be heated up simultaneous, as reflected in case 2. The results indicate that much more net extra gas can be produced when heat transfer efficiency is improved, even within 5 years of thermal stimulation. It also shows that longer stimulation time leads to more net recovery of extra gas in absolute terms, but result in less production of extra gas by percentage. This implies that if we assess the thermal stimulation efficiency using net extra gas produced by percentage as an indicator, there exists an optimal duration on how long thermal stimulation should be applied.

|  |  | Case 1 |  | Case 2 |
|---|---|---|---|---|
|  | Matrix initial permeability (nd) | 10 | 100 | 100 |
| 5$^{th}$ year | Energy input (billion J) | 820000 | 820000 | 1860000 |
|  | Equivalent gas consumption (billion m^3) | 0.021 | 0.021 | 0.048 |
|  | Net extra gas produced (billion m^3) | 0.017 | 0.021 | 0.097 |
|  | Net extra gas produced by percentage | 18.78% | 14.95% | 69.95% |
| 15$^{th}$ year | Energy input (billion J) | 1320000 | 1320000 | 2000000 |
|  | Equivalent gas consumption (billion m^3) | 0.034 | 0.034 | 0.052 |
|  | Net extra gas produced (billion m^3) | 0.040 | 0.042 | 0.115 |
|  | Net extra gas produced by percentage | 30.01% | 22.25% | 61.44% |

**Table 2. Energy balance calculation at the 5$^{th}$ and 15$^{th}$ benchmark years**

From the above results and analysis, we can conclude that thermal stimulation has great potential to improve shale gas ultimate recovery significantly by allowing the release of residual adsorption gas that cannot be recovered by pressure depletion alone, and increasing the overall flow capacity in the formation, by increasing matrix apparent permeability and relaxing/delaying the reduction of fracture conductivity. However, the effectiveness of thermal stimulation on ultimate recovery largely hinges on how much time it takes to elevate the formation temperature to a target level. Because the elevation of formation temperature has compound effects on all the coupled underlying physics, and the impact of these effects are cumulative. The sooner the formation can be heated up to the desired temperature, the longer time the reservoir can benefit from its altered properties, and hence, the effectiveness of thermal stimulation can be enhanced significantly.

## Conclusions and Discussions

In this study, a thermal stimulation method that combines transverse hydraulic fracture and electromagnetically sensitive proppants is proposed, and the possibility of enhancing shale gas recovery behavior by elevating formation temperature is explored. In order to assess the overall effects of thermal stimulation on gas production, we presented general formulations of numerically modeling and simulation of gas production in fractured shale gas reservoirs for the first time, with fully coupled thermal-hydraulic-mechanical mechanisms. The process of real gas flow in shale matrix/ fractures, rock deformation due to changing in-situ stress and heat transfer by conduction are coupled together through temperature-pressure dependent variables. The results indicate that the gas production and ultimate recovery have the potential to be improved substantially by elevating temperature, which alters the gas adsorption/desorption behavior and the overall flow capacity in fractured shales. But the heat transfer efficiency (i.e., how long it will take to heat up the drainage volume to desired temperature) has dominant impact on the effectiveness of thermal stimulation.

It should be mentioned that this paper only focuses on the effects thermal stimulation on gas adsorption/desorption behavior and the evolution of matrix permeability and fracture conductivity during production. It would be more realistic if the shear-induced slippage, rock failure due to thermal stress, and all the other possible mechanisms (such as the impact of temperature on clay bound water, capillary bound water and the transformation of organic matter into hydrocarbons) can be incorporated into the model to assess the overall thermal stimulation effects. In addition, the applications of other thermal stimulation methods, such as microwave, nano-particle fracturing fluid, etc., need to be explored to improve the efficiency of heat transfer process. Because different shale plays have different rock, mineral, and organic matter compositions, their gas adsorption capacity have various sensitivity to temperature, so laboratory experiment is crucial in understanding to what degree the temperature can impact gas adsorption/desorption process and how to optimize thermal stimulation design correspondingly. Considering an average shale gas recovery factor of 30% in U.S. shale plays, there is a big prize to be claimed in terms of enhanced recovery using thermal stimulation techniques in shale gas reservoirs, however, not much work has been done in this area yet. The general approaches and the proposed mathematical framework of thermal-hydraulic-mechanical (THM) modeling presented in this article set a foundation for future research efforts, to better understand fully-coupled physics and explore new enhanced recovery techniques in fractured shale gas reservoirs.

## Appendix
## Mathematical Formulations of Thermal-Hydraulic-Mechanical Modeling in Fractured Shale Gas Reservoirs

The fully coupled THM physical process in shale gas reservoirs involves fluid flow within the formation matrix and fractures, shale gas adsorption and desorption, real gas properties and in-situ stress that affected by both pressure and temperature locally. In this section, the general governing equations of the coupled fluid flow, geomechanics and heat transfer process will be presented first, and then the pressure-temperature dependent variables will be discussed separately. All the equations presented in this section do not have any unit conversion parameters, any unit system can be applied (e.g., SI unit or Oil Field unit), as long as they are consistent.

### Gas Flow in Formation Matrix and Fractures

Due to the extremely small pore radii in shale gas formations, Non-Darcy flow/Gas-Slippage can have a significant impact on gas transport in porous medium. Ideally, Non-Darcy flow mechanism can be modeled accurately using molecular physics by capturing the interaction between molecules and pore walls on nanoscale, however, this technique is not practical for modeling flow through shale on reservoirs scale. To make simulation feasible, it is desirable to integrate molecular flow behavior with the standard Darcy's equation so that these mechanisms can be captured without extremely intensive computational efforts. Non-Darcy flow based on matrix apparent permeability can be used to describe the gas flow rate within the shale matrix:

$$\boldsymbol{q}_g = -\frac{\boldsymbol{k}_a}{\mu_g} \cdot \nabla P \tag{A.1}$$

where $\boldsymbol{q}_g$ is the gas velocity vector, $\mu_g$ is the gas viscosity, and $\boldsymbol{k}_a$ is the matrix apparent permeability tensor, which is pressure and temperature dependent. The continuity equation within shale gas formation can be written as:

$$\frac{\partial m}{\partial t} + \nabla \cdot (\rho_g \boldsymbol{q}_g) = Q_m \tag{A.2}$$

where $m$ is the gas content per unit volume, $\rho_g$ is gas density, $Q_m$ is the source term and $t$ is the generic time. The gas content $m$ is obtained from two contributions:

$$m = \rho_g \emptyset_m + m_{ad} \tag{A.3}$$

where $\emptyset_m$ is matrix porosity, $\rho_g \emptyset_m$ is the free gas mass in the shale pore space per unit volume of formation, while $m_{ad}$ is the adsorbed gas mass per unit volume of formation.

Normally, gas flow rate inside fractures in shale reservoirs are much smaller when compared to the conventional reservoir, due to extremely small shale matrix permeability, and turbulence flow is unlikely to happen under this circumstances, so Darcy's law is adequate enough to describe the flow behavior inside the fractures. Discrete Fracture Network (DFN) with tangential derivatives can be used to define the flow along the interior boundary representing fractures within the porous medium.

$$\boldsymbol{q}_f = -\frac{\boldsymbol{k}_f}{\mu_g} \cdot d_f \nabla_T P \tag{A.4}$$

where $\boldsymbol{q}_f$ is the gas volumetric flow rate vector per unit length in the fracture, $\boldsymbol{k}_f$ is the fracture permeability tensor, $d_f$ is fracture width and $\nabla_T P$ is the pressure gradient tangent to the fracture surface. The continuity equation along the fracture reflects the generic material balance within the fracture:

$$d_f \frac{\partial \emptyset_f \rho_g}{\partial t} + \nabla_T \cdot (\rho_g \boldsymbol{q}_f) = d_f Q_f \tag{A.5}$$

where $\emptyset_f$ is the fracture porosity, and $Q_f$ is the mass source term, which can be calculated by adding the mass flow rate per unit volume from two fracture walls:

$$Q_f = Q_{left}^f + Q_{right}^f \tag{A.6-a}$$

$$Q_{left}^f = -\frac{\boldsymbol{k}_a}{\mu_g} \frac{\partial P_{left}}{\partial \boldsymbol{n}_{left}} \tag{A.6-b}$$

$$Q_{right}^f = -\frac{\boldsymbol{k}_a}{\mu_g} \frac{\partial P_{right}}{\partial \boldsymbol{n}_{right}} \tag{A.6-c}$$

where $\boldsymbol{n}$ is the vector perpendicular to fracture surface.

DFN is treated as an internal boundary condition, and it has one less dimension than the simulation domain (e.g., if reservoir domain is 2D, DFN is modeled as 1D internal line-boundary; if reservoir domain is 3D, DFN is modeled as 2D internal face-boundary). This reflects the reality for fractures with small apertures, where flow in the width direction within the fracture is negligible.

**Stress Equilibrium**

The porous medium is assumed to be perfectly elastic so that no plastic deformation occurs. The constitutive equation can be expressed in terms of effective stress ($\sigma_{ij}$,) strain ($\varepsilon_{ij}$), pore pressure (P) and temperature (T):

$$\sigma_{ij} = 2G\varepsilon_{ij} + \frac{2G\nu}{1-2\nu}\varepsilon_{kk}\delta_{i,j} - \alpha P\delta_{i,j} + \beta\Delta TE\delta_{i,j} \tag{A.7}$$

where G is the shear modulus, E is the Young's modulus, $\nu$ is the Poisson's ratio, $\varepsilon_{kk}$ represents the volumetric strain, $\delta_{i,j}$ is the Kronecker delta defined as 1 for $i = j$ and 0 for $i \neq j$, $\beta$ is thermal expansion coefficient and $\alpha$ is the Biot's effective stress coefficient, which is assumed to be 1 in this study. The strain-displacement relationship and equation of equilibrium are defined as:

$$\varepsilon_{ij} = \frac{1}{2}(u_{i,j} + u_{j,i}) \tag{A.8}$$
$$\sigma_{ij,j} + F_i = 0 \tag{A.9}$$

where $u_i$ and $F_i$ are the components of displacement and net body force in the $i$-direction. Combining Eq. A.7– A.9, we have a modified Navier equation in terms of displacement under a combination of applied stress, pore pressure and temperature variations:

$$G\nabla u_i + \frac{G}{1-2\nu}u_{j,ii} - \alpha P\delta_{i,j} + \beta\Delta TE\delta_{i,j} + F_i = 0 \tag{A.10}$$

**Heat Transfer in Porous Medium**

Heat transfer process is governed by a thermal diffusion equation in an isotropic porous medium and the radiative effects, viscous dissipation, and work done by pressure changes are negligible. Considering an elemental volume of a porous medium we have, for the matrix:

$$(1-\emptyset_m)\rho_m C_m \frac{\partial T_m}{\partial t} + (1-\emptyset_m)\nabla \cdot (-\lambda_m \nabla T_m) = (1-\emptyset_m)h_m \tag{A.11}$$

and for the gas phase:

$$\emptyset_m \rho_g C_{p,g} \frac{\partial T_g}{\partial t} + \emptyset_m \nabla \cdot (-\lambda_g \nabla T_g) + \rho_g C_{p,g} \boldsymbol{q}_g \cdot \nabla T_g = \emptyset_m h_g \tag{A.12}$$

where $C_m$ is the heat capacity of the rock matrix, $C_{p,g}$ is heat capacity at constant pressure of the gas phase, $\lambda$ is thermal conductivity, $h$ is the heat source term, $\rho$ is the density and the subscripts m and g refer to formation matrix and gas phases respectively. Setting $T_m = T_g = T$ by assuming there is local thermal equilibrium at the wall of pores, and adding Eq.( A.11) and (A.12) we have

$$(\rho C_p)_{eq} \frac{\partial T}{\partial t} + \nabla \cdot (-\lambda_{eq} \nabla T) + \rho_g C_{p,g} \boldsymbol{q}_g \cdot \nabla T = h_{eq} \tag{A.13 − 1}$$

where

$$(\rho C_p)_{eq} = (1-\emptyset_m)\rho_m C_m + \emptyset_m \rho_g C_{p,g} \tag{A.13 − 2}$$

$$\lambda_{eq} = (1-\emptyset_m)\lambda_m + \emptyset_m \lambda_g \tag{A.13 − 3}$$

$$h_{eq} = (1-\emptyset_m)h_m + \emptyset_m h_g \tag{A.13 − 4}$$

$(\rho C_p)_{eq}$ is the equivalent volumetric heat capacity at constant pressure, $\lambda_{eq}$ is the equivalent thermal conductivity and $h_{eq}$ is the overall equivalent heat source. Typically, the heat capacity of most formation rock is at least hundred times larger than that of gas phase and formation porosity is normally less than 3% in shale formation, so $(\rho C_p)_{eq}$, $\lambda_{eq}$ and $h_{eq}$ are dominated by the term of $(1-\emptyset_m)\rho_m C_m$, $(1-\emptyset_m)\lambda_m$ and $(1-\emptyset_m)h_m$, respectively. In addition, the gas flow rate is constrained by extremely low matrix permeability and with low gas density, the term $\rho_g C_{p,g}\boldsymbol{q}_g \cdot \nabla T$ diminishes, so the influence of heat transfer by convection is negligible, and the heat conduction mechanism in the formation matrix dominates the entire heat transfer process. Consequently Eq. (A.13-1) can be simplified as the following:

$$\rho_m C_m \frac{\partial T}{\partial t} + \nabla \cdot (-\lambda_m \nabla T) = h_{eq} \tag{A.14}$$

Overall, with appropriate initial and boundary conditions, Eq. (A.2), (A.5), (A.10) and (A.14) can well describe the fully coupled THM system though the pressure and temperature dependent variables in those equations.

**Pressure-Temperature Dependent Variables**

In the following sections, the pressure and temperature dependent variables that needed for coupling process in the above governing equations will be determined, using either theoretical modeling or well-established correlations.

**Real Gas Properties**

The in-situ gas density can be calculated by real gas law:

$$\rho_g = \frac{PM}{ZRT} \tag{A.15}$$

where $M$ is the average molecular weight of mixed gas, R is the universal gas constant. The Z-factor can be estimated by solving Equation of State (EOS) or using correlations for the gas mixtures. In this study, the Z-factor is calculated using an explicit correlation (Mahmoud, 2013) based on the pseudo-reduced pressure ($p_{pr}$) and pseudo-reduced temperature ($T_{pr}$):

$$Z = (0.702 e^{-2.5 T_{pr}})(p_{pr}^2) - (5.524 e^{-2.5 T_{pr}})(p_{pr}) + (0.044 T_{pr}^2 - 0.164 T_{pr} + 1.15) \tag{A.16}$$

The advantage of using explicit correlation is to avoid solving higher order equations respect to Z-factor, which leads to multiple solutions and increases computation efforts.

Among all gas properties, gas viscosity, which is usually characterized by some available correlations, still remains uncertain. Lee et al. (1966) proposed a correlation to calculate gas viscosity at temperatures from 310 to 445 K and pressure from 0.69 to 55.16 MPa. Viswanathan (2007) measured the viscosity of pure methane at pressure from 34.47 to 206.84 MPa and temperatures from 310 to 478 K, and modified the gas viscosity correlation by Lee et al. (1966):

$$\mu_g = 10^{-4} K e^{X \rho_g^Y} \tag{A.17-1}$$

where

$$K = \frac{(5.0512 - 0.2888 M) T^{1.832}}{-443.8 + 12.9 M + T} \tag{A.17-2}$$

$$X = -6.1166 + \left(\frac{3084.9437}{T}\right) + 0.3938 M \tag{A.17-3}$$

$$Y = 0.5893 + 0.1563 X \tag{A.17-4}$$

**Gas Adsorption Capacity**

Langmuir isotherms (1916) are widely used to model adsorption gas content in shales as a function of pressure, which can be established for a prospective area of shale basin using available data on TOC and on thermal maturity to establish the Langmuir volume ($V_L$) and the Langmuir pressure ($P_L$). And adsorbed gas-in-place is then calculated using the formula below:

$$m_{ad} = \rho_m \rho_{gst} V_L \frac{P}{P + P_L} \tag{A.18}$$

where $\rho_{gst}$ is gas density at standard condition. However, the temperature effect on gas adsorption capacity is not included in Langmuir isotherms. Lu et al. (1995) investigated temperature dependent adsorption curves on shale samples and proposed a Bi-Langmuir model that accounts for gas adsorption on both clay minerals and kerogen:

$$m_{ad} = \rho_m \rho_{gst} V_L \left( f_1 \frac{K_1(T) P}{1 + K_1(T) P} + f_2 \frac{K_2(T) P}{1 + K_2(T) P} \right) \tag{A.19-1}$$

where $f_i$ is defined as the ratio of the amount of the ith type of adsorption at monolayer coverage to the total amount adsorbed at monolayer coverage and each adsorption site is assumed to follow the Langmuir equation. For two types of adsorption sites (clay minerals and kerogen) we have:

$$f_1 + f_2 = 1 \tag{A.19-2}$$

$$K_1(T) = a_1 T^{-\frac{1}{2}} e^{-\frac{J_1}{RT}} \qquad (A.19-3)$$

$$K_2(T) = a_2 T^{-\frac{1}{2}} e^{-\frac{J_2}{RT}} \qquad (A.19-4)$$

where $a_1$ and $a_2$ is a pre-exponential constant independent of temperature, $J_1$ and $J_2$ are the characteristic adsorption energy. The five unknown independent parameters $f_1$, $a_1$, $a_2$, $J_1$ and $J_2$ in Eq.( A.19) can be determined from a Langmuir isotherm curve at provided temperature condition by non-linear regression. Thus, the temperature effects can be included to describe shale gas adsorption capacity as a function of both pressure and temperature (as shown in Fig.6), with available Langmuir volume and Langmuir pressure values as input parameters (Wang et al., 2014), or directly fitting and validated against experiment data within possible temperature ranges (Yue et al., 2015).

**Fracture Pressure Dependent Permeability**

It is a well-known fact that most shale formations have massive pre-existing natural fracture networks that are generally sealed by precipitated materials weakly bonded with mineralization. Such poorly sealed natural fractures are generally reported to interact heavily with the hydraulic fractures during the injection treatments, serving as preferential paths for the growth of complex fracture network. The reactivation of such pre-existing planes of weakness (i.e., natural fractures, micro-fractures, fissures) is well documented and observed in microseismic monitoring (Cipolla et al. 2011; Zakhour et al. 2015). Ideally, the fracture conductivity on un-propped and propped fracture surfaces, as a function of confining stress, fluid type, and proppant type should be investigated for a specific reservoir in order to optimize stimulation design and field development, and the fracture pressure-dependent permeability (PDP) data should be incorporated into reservoir simulation model to better forecast the well performance under various operation conditions. The expected reduction in fracture permeability caused by the increasing effective stress during production is can be described by power law relationship (Cho et al. 2013), here, we use the following correlation:

$$k_f = k_{f,i} e^{-B\sigma_m} \qquad (A.20)$$

where $k_f$ is the fracture permeability, $k_{f,i}$ is the fracture permeability at initial reservoir conditions, B is a fracture compaction parameter that can be determined from experimental data, and $\sigma_m$ is the mean effective stress, which is influenced by declining pressure and rising temperature during stimulation.

**Correction for Gas Apparent Permeability in Nano-Pores**

Darcy's law cannot describe the actual gas behavior and transport phenomena in nano-porous media. In such nanopore structure, fluid flow departs from the well-understood continuum regime, in favor of other mechanisms such as slip, transition, and free molecular conditions. The Knudsen number (Knudsen, 1909) is a dimensionless parameter that can be used to differentiate flow regimes in conduits at micro and nanoscale. For conduit with radius r, Knudsen number can be estimated by

$$K_n = \frac{\mu_g Z}{Pr} \sqrt{\frac{\pi RT}{2M}} \qquad (A.21)$$

**Table A1** shows how these different flow regimes, which correspond to specific flow mechanisms, can be classified by different ranges of $K_n$.

| $K_n$ | $0 - 10^{-3}$ | $10^{-3} - 10^{-1}$ | $10^{-1} - 10^1$ | $> 10^1$ |
|---|---|---|---|---|
| Flow Regime | Continuum | Slip | Transition | Free Molecular |

**Table A1. Fluid Flow Regimes Defined by Ranges of $K_n$ (Roy et al. 2003)**

The apparent permeability of shale matrix can be represented by the following general form:

$$k_a = k_\infty f(K_n) \qquad (A.22)$$

where $k_\infty$ is the intrinsic permeability of the porous medium, which is defined as the permeability for a viscous, non-reacting ideal liquid, and it is determined by the nano-pore structure of porous medium itself. $f(K_n)$ is the correlation term that relates the matrix apparent permeability and intrinsic permeability. Sakhaee-Pour and Bryant (2012) developed correlations which are based on lab experiments. They proposed a first-order permeability model in the slip regime and a polynomial form for the permeability enhancement in the transition regime using regression method:

$$f(K_n) = \begin{cases} 1 + \alpha_1 K_n & \text{Slip Regime} \\ 0.8453 + 5.4576 K_n + 0.1633 K_n^2 & \text{Transition Regime } 0.1 < K_n < 0.8 \end{cases} \qquad (A.23)$$

where $\alpha_1$ is permeability enhancement coefficient in slip regime. To ensure the approximation of continuity of $f(K_n)$ at the boundary region of slip and transition regime, where no existing model available, $\alpha_1$ is set to be 4. For all the simulation cases presented in this study, $K_n$ is always larger than 0.1, so only the polynomial expression for transition regime is automatically used in our simulator.

Intuitively, Eqs. (A.21) and (A.23) predict net increases of $K_n$ and $f(K_n)$ with decreasing pore pressure, respectively. However, decreasing pore pressure can also lead to reduction in pore radii and, in turn, reduce the intrinsic permeability $k_\infty$. It can be seen from Eq. (A.22) that the evolution of matrix apparent permeability $k_a$ is determined by two combined mechanisms (i.e., the variations in $k_\infty$ and $f(K_n)$ during production).

Even though many studies (Lunati and Lee 2014; Sigal and Qin 2008; Kazemi and Takbiri-Borujeni 2015) have developed various theoretical models to account for the effects of permeability enhancement in nano-pore structure of shale gas, but how to quantify these effects in complex, heterogeneous shale systems on a field scale is still not well understood within scientific and industry community. In addition, all these proposed models do not account for the dynamic changes of stress and adsorption layer as pressure declines (as reflected in **Fig.2**) in a fractured system, which can significantly overestimate the effects of permeability enhancement on production and production decline trend (Wang 2016c). Recent study (Wang et al 2015b) also reveals that under most real shale-gas-reservoir conditions, gas adsorption and the non-Darcy flow are dominant mechanisms in affecting apparent permeability, while the effects of surface diffusion caused by adsorbed gas can be largely overlooked. So in this article, molecule diffusion will not be discussed in the following derivations for matrix apparent permeability.

Wang and Marongiu-Porcu (2015) conducted a comprehensive literature review on gas flow behavior in shale nano-pore space and proposed a unified matrix apparent permeability model, which bridges the effects of geomechanics, non-Darcy flow and gas adsorption layer into a single mode, by considering the microstructure changes in nano-pore space. In a general porous media, the loss in cross-section area is equivalent to the loss of porosity,

$$\frac{\phi}{\phi_0} = \frac{r^2}{r_0^2} \tag{A.24}$$

where variables with a subscript 0 correspond to their value at the reference state, which can be laboratory or initial reservoir conditions. Laboratory measurements by Dong et al. (2010) shows that the relationship between porosity and stress follows a power law relationship, and can be expressed using the concept of mean effective stress $\sigma_m$ (Wang and Marongiu-Porcu 2015):

$$\phi = \phi_0 \left(\frac{\sigma_m}{\sigma_{m0}}\right)^{-C_\phi} \tag{A.25}$$

$C_\phi$ is a dimensionless material-specific constant that can be determined by lab experiments. For silty-shale samples, the values of $C_\phi$ range from 0.014 to 0.056 (Dong et al., 2010). Combining Eq. (A.24) and Eq. (A.25) leads to the relationship between pore radius and local stress

$$r = r_0 \left(\frac{\sigma_m}{\sigma_{m0}}\right)^{-0.5 C_\phi} \tag{A.26}$$

When the adsorption layer is considered, the thickness of the gas adsorption layer, $\delta$, can be interpolated based on a Langmuir type functional relationship:

$$\delta = d_m \frac{P/P_L}{1 + P/P_L} \tag{A.27}$$

where $d_m$ is the average diameter of gas molecules residing on the pore surface. And the effective pore radius (A.26) can be modified as:

$$r = r_0 \left(\frac{\sigma_m}{\sigma_{m0}}\right)^{-0.5 C_\phi} - d_m \frac{P/P_L}{1 + P/P_L} \tag{A.28}$$

The relationship between intrinsic permeability and pore radius can have the following relationship

$$\frac{k_\infty}{k_{\infty 0}} = \left(\frac{r}{r_0}\right)^\eta \tag{A.29}$$

where $\eta$ is the coefficient that define the sensitivity of permeability to the changes of pore radius. Different shale formations may have different nano-pore structure typology, which leads to different values of $\beta$. In this study, $\eta$ equals 2 by assuming the overall all intrinsic permeability resembles fluid flow in a capillary tube (Beskok and Karniadakis 1999). Combining Eq. (A.28), Eq. (A.29) and Eq. (A.22), we have the final expression of matrix apparent permeability:

$$k_a = k_{\infty 0} \frac{\left(r_0 \left(\frac{\sigma_m}{\sigma_{m0}}\right)^{-0.5 C_\emptyset} - d_m \frac{P/P_L}{1 + P/P_L}\right)^\eta}{r_0^\eta} f(K_n) \tag{A.30}$$

Compared to the apparent matrix permeability model proposed by Wang and Marongiu-Porcu (2015), Eq. (**A.30**) is a more general formula for shale matrix apparent permeability, with additional parameter **η** to relate the changes in pore radius to the alterations in intrinsic permeability. **η** can be determined from laboratory experiment for different types of shales.

As the pore pressure declines and temperature increases, formation porosity, pore radius, Knudsen number and intrinsic permeability, are subject to time and space variation in SRV, and so does the matrix apparent permeability. This matrix apparent permeability model enables us to assess the evolution of matrix flow capacity under coupled effects of pressure depletion and heat transfer process during thermal stimulation. With all these pressure and temperature dependent variables substituting into the general governing equations described in the previous sections, a closed system of fully coupled thermal-hydraulic-mechanic shale gas reservoir simulation model is constructed and ready to be solved by available numerical methods. Based on the formulas introduced above, a unified shale gas reservoir simulator was constructed. Newton-Raphson method (Ypma, 1995) and finite element analysis (Zienkiewicz and Taylor, 2005) are used to solve all the coupled equations numerically.

**Nomenclature**

| | |
|---|---|
| $a_i$ | = Pre-exponential constant for ith type of adsorption, $Lt^2/m$, $1/Pa$ |
| $B$ | = Parameter for fracture pressure dependent permeability, $Lt^2/m$, $1/Pa$ |
| $C_m$ | = Heat capacity of rock matrix, $L^2/t^2/T$, J/K/kg |
| $C_{p,g}$ | = Heat capacity of gas phase at constant pressure, $L^2/t^2/T$, J/K/kg |
| $C_\emptyset$ | = Material constant for pressure dependent porosity |
| $d_f$ | = Fracture width, L, $m$ |
| $d_m$ | = Diameter of absorbed gas molecules, L, $m$ |
| $E$ | = Young's modulus, $m/Lt^2$, $Pa$ |
| $f(K_n)$ | = Non-Darcy flow correction term |
| $f_i$ | = Fraction of ith type of adsorption |
| $F_i$ | = Net body force along i direction, $m/Lt^2$, $Pa$ |
| $G$ | = Shear modulus, $m/Lt^2$, $Pa$ |
| $h_{eq}$ | = Equivalent heat source term, $m/t^3/L$, $J/m^3/s$ |
| $h_g$ | = Heat source term in gas phase, $m/t^3/L$, $J/m^3/s$ |
| $h_m$ | = Heat source term in rock matrix, $m/t^3/L$, $J/m^3/s$ |
| $J_i$ | = Characteristic adsorption energy for ith type of adsorption, $mL^2/t^2/n$, J/mol |
| $k_a$ | = Apparent gas permeability, $L^2$, $m^2$ |
| $\boldsymbol{k_a}$ | = Gas apparent permeability tensor, $L^2$, $m^2$ |
| $k_f$ | = Fracture permeability, $L^2$, $m^2$ |
| $\boldsymbol{k_f}$ | = Fracture permeability tensor, $L^2$, $m^2$ |
| $k_{f,i}$ | = Fracture permeability at initial reservoir conditions, $L^2$, $m^2$ |
| $k_\infty$ | = Matrix intrinsic permeability, $L^2$, $m^2$ |
| $k_{\infty 0}$ | = Matrix intrinsic permeability at reference conditions, $L^2$, $m^2$ |
| $K_n$ | = Knudsen number |
| $L$ | = Characteristic length of flow path, L, $m$ |
| $m$ | = Total gas content, $m/L^3$, $kg/m^3$ |
| $m_{ad}$ | = Gas adsorption mass per unit volume, $m/L^3$, $kg/m^3$ |
| $M$ | = Molecular weight, $m/n$, $kg/mol$ |
| $\boldsymbol{n}$ | = Normal vector to fracture surface |
| $P$ | = Reservoir pressure, $m/Lt^2$, $Pa$ |
| $P_e$ | = Effective confined pressure, $m/Lt^2$, $Pa$ |
| $P_L$ | = Langmuir pressure, $m/Lt^2$, $Pa$ |

| $P_{pr}$ | = Pseudo-reduced pressure |
| $P_0$ | = Pressure at reference condition, m/L$t^2$, $Pa$ |
| $\mathbf{q}_f$ | = Flow rate vector in the fracture per unit length, $L^2$/t, $m^3/s/m$ |
| $\mathbf{q}_g$ | = Velocity vector of gas phase, L/t, $m/s$ |
| $Q_f$ | = Mass source term in fracture, m/$L^3$/t, $kg/m^3/s$ |
| $Q_m$ | = Mass source term in matrix, m/$L^3$/t, $kg/m^3/s$ |
| $r$ | = Effective pore radius, L, $m$ |
| $r_0$ | = Effective pore radius at reference conditions, L, $m$ |
| $R$ | = Universal gas constant, m$L^2/t^2/n/T$, 8.3145 $J/mol/K$ |
| $T$ | = Reservoir temperature, T, $K$ |
| $T_g$ | = Temperature of gas phase, T, K |
| $T_m$ | = Temperature of rock matrix, T, K |
| $T_{pr}$ | = Pseudo-reduced temperature |
| $u_{i,j}$ | = Component of displacement, L, m |
| $V_L$ | = Langmuir volume, $L^3$/m, $m^3/kg$ |
| $Z$ | = Gas deviation factor |
| $\alpha$ | = Biot's coefficient |
| $\alpha_1$ | = permeability enhancement coefficient |
| $\beta$ | = Thermal expansion coefficient, 1/T, 1/K |
| $\delta$ | = Thickness of gas adsorption layer, L, $m$ |
| $\delta_{i,j}$ | = Kronecker delta |
| $\varepsilon_{ij}$ | = Elastic strain |
| $\varepsilon_{kk}$ | = Volumetric strain |
| $\eta$ | = Coefficient for pore radius dependent intrinsic permeability |
| $\lambda_{eq}$ | = Equivalent thermal conductivity, m$L/t^3/T$, W/m/K |
| $\lambda_g$ | = Thermal conductivity of gas phase, m$L/t^3/T$, W/m/K |
| $\lambda_m$ | = Thermal conductivity of rock matrix, m$L/t^3/T$, W/m/K |
| $\mu_g$ | = Gas viscosity, m/Lt, Pa s |
| $\nu$ | = Poisson's ratio |
| $\rho_g$ | = Gas density, m/$L^3$, $kg/m^3$ |
| $\rho_{gst}$ | = Gas density at standard condition, m/$L^3$, $kg/m^3$ |
| $\rho_m$ | = Matrix density, m/$L^3$, $kg/m^3$ |
| $(\rho C_p)_{eq}$ | = Equivalent volumetric heat capacity, m/$t^2$/T/L, J/K/$m^3$ |
| $\sigma_{ij}$ | = Effective stress, m/L$t^2$, $Pa$ |
| $\sigma_m$ | = Mean effective stress, m/L$t^2$, $Pa$ |
| $\sigma_{m0}$ | = Mean effective stress at reference conditions, m/L$t^2$, $Pa$ |
| $\phi_f$ | = Fracture porosity |
| $\phi_m$ | = Matrix in situ porosity |
| $\phi_{m0}$ | = Matrix porosity at reference conditions |

**References**

Ahmed Elfeel, M., Jamal, S., Enemanna, C., Arnold, D., and Geiger, S. 2013. Effect of DFN Upscaling on History Matching and Prediction of Naturally Fractured Reservoirs. SPE Paper 164838 presented at the EAGE Annual Conference & Exhibition, held in London, UK, 10-13 June. http://dx.doi.org/10.2118/164838-MS

Aimene, Y and Ouenes, A. 2015. Geomechanical modeling of hydraulic fractures interacting with natural fractures — Validation with microseismic and tracer data from the Marcellus and Eagle Ford. Interpretation, 3(3): SU71-SU88. http://dx.doi.org/ 10.1190/INT-2014-0274.1

Ambrose, R.J., Hartman, R.C. and Diaz-Campos, M. 2010. New pore-scale considerations for shale gas in place calculations. In: SPE Unconventional Gas Conference, 23-25 February, Pittsburgh, Pennsylvania, USA. http://dx.doi.org/10.2118/131772-MS

Burghignoli, A., A. Desideri, and S. Miliziano, A laboratory study on the thermomechanical behaviour of clayey soils. Canadian Geotechnical Journal, 2000. 37(4): p. 764–780. http://dx.doi.org/10.1139/t00-010

Cekerevac, C. and L. Laloui, Experimental Study of Thermal Effects on the Mechanical Behaviour of a Clay. International Journal for Numerical and Analytical Methods in Geomechanics, 2004. 28(3): p. 209-228. http://dx.doi.org/10.1002/nag.332

Chapiro, G. C., and Bruining, J. 2014. Thermal Well Stimulation in Gas Shales Through Oxygen Injection and Combustion. Fourth EAGE Shale Workshop. http://doi.org/10.3997/2214-4609.20140038

Checkhonin, A., Parshin, A., Pissarenko, D., Popov, Yury., Romushkevich, R., Safonov, S., Spasennykh, M., Chertenkov, M., and Stenin, V. 2012. When Rocks Get Hot: Thermal Properties of Reservoir Rocks. Oilfield Review 24, no. 3. 20-37

Cho, Y., Ozkan, E., and Apaydin, O. G. 2013. Pressure-Dependent Natural-Fracture Permeability in Shale and Its Effect on Shale-Gas Well Production. SPE Reservoir Evaluation & Engineering, **16**(02), 216–228. http://dx.doi.org/10.2118/159801-PA

Cipolla, C. L., Mack, M. G., Maxwell, S. C., and Downie, R. C. 2011. A Practical Guide to Interpreting Microseismic Measurements. Paper SPE 144067 presented at the North American Unconventional Gas Conference and Exhibition, The Woodlands, Texas 14–16 June. http://dx.doi.org/10.2118/144067-MS

Cornette, B. M., Telker, C., and De La Pena, A. 2012. Refining Discrete Fracture Networks With Surface Microseismic Mechanism Inversion and Mechanism-Driven Event Location. SPE Paper 151964 presented at the SPE Hydraulic Fracturing Technology Conference, held in The Woodlands, Texas, USA, 6-8 February. http://dx.doi.org/10.2118/151964-MS

Darishchev, A., Rouvroy, P. and Lemouzy, P. 2013. On Simulation of Flow in Tight and Shale Gas Reservoirs. Paper was presented at the SPE Unconventional Gas Conference and Exhibition, held at Muscat, Oman, January 28-30. http://dx.doi.org/10.2118/163990-MS

Dong, J.J., Hsu, J.Y., and Wu, W.J. 2010. Stress-Dependence of the Permeability and Porosity of Sandstone and Shale from TCDP Hole-A. International Journal of Rock Mechanics and Mining Sciences **47** (7):1141–1157. http://dx.doi.org/10.1016/j.ijrmms.2010.06.019

Fathi, E., Tinni, A., and Akkutlu I.Y. 2012. Shale Gas Correction to Klinkenberg Slip Theory. Paper SPE 154977 presented at the SPE Americas Unconventional Resources Conferences at Pittsburgh, Pennsylvania 5–7 June. http://dx.doi.org/10.2118/154977-MS

Ghassemi, A., and Suarez-Rivera, R. 2012. Sustaining Fracture Area and Conductivity of Gas Shale Reservoirs for Enhancing Long-Term Production and Recovery. RPSEA Report. http://www.rpsea.org/projects/08122-48

Gregg, S.J., Sing, K.S.W. 1982. Adsorption, surface area and porosity, 2nd edition. Academic Press, New York, Page 303. http://doi.org/10.1524/zpch.1969.63.1_4.220

Guo, J.C., Zhao, X., Zhu, H.Y., Zhang, X.D., Pan, R., 2015. Numerical simulation of interaction of hydraulic fracture and natural fracture based on the cohesive zone finite element method. Journal of Natural Gas Science and Engineering. 25, 180-188. http://dx.doi.org/ 10.1016/j.jngse.2015.05.008

Hascakir, B. 2008. Investigation of productivity of heavy oil carbonates and oil shales using electrical heating methods. PhD dissertation, Middle East Technical University, Ankara, Turkey.

Hoda, N., Fang, C., Lin, M., Symington, W and Stone, M. 2010. Numerical Modeling of ExxonMobil's Electrofrac Field Experiment at Colony Mine, Paper presented at the 30[th] Oil Shale Symposium, held in Denver, Colorado, 18-20 October.

Johri, M., and Zoback, M. D. 2013. The Evolution of Stimulated Reservoir Volume during Hydraulic Stimulation of Shale Gas Formations. Paper SPE presented at the Unconventional Resources Technology Conference, Denver, Colorado, USA, 12-14 August. http://dx.doi.org/10.1190/URTEC2013-170

Johri, M., and Zoback, M. D. 2013. The Evolution of Stimulated Reservoir Volume during Hydraulic Stimulation of Shale Gas Formations. Paper SPE 168701 presented at the Unconventional Resources Technology Conference, Denver, Colorado, 12-14 August, http://dx.doi.org/10.1190/URTEC2013-170

Kazemi, M and Takbiri-Borujeni,A. 2015. An analytical model for shale gas permeability, International Journal of Coal Geology 146:188-197. http://dx.doi.org/10.1016/j.coal.2015.05.010

Knudsen, M. 1909. Die Gesetze der Molekularströmung und der inneren Reibungsströmung der Gase durch Röhren (The laws of molecular and viscous flow of gases through tubes). Ann. Phys. **333**(1): 75–130. http://dx.doi.org/10.1002/andp.19093330106

Laloui, L. and C. Cekerevac, Thermo-Plasticity of Clays: An Isotropic Yield Mechanism. Computers and Geotechnics, 2003. 30(8): p. 649-660. http://doi.org/10.1016/j.compgeo.2003.09.001

Langmuir, I. 1916. The Constitution and Fundamental Properties of Solids and Liquids. Journal of the American Chemical Society, **38**(11):2221–2295. http://dx.doi.org/ 10.1021/ja02268a002

Lee, A.L., Gonzalez, M.H., and Eakin, B.E. 1966. The Viscosity of Natural Gases. JPT 997; Trans., AIME, 237. http://dx.doi.org/10.2118/1340-PA

Lu, X.C., Li, F.C. and Watson, A.T. 1995. Adsorption Studies of Natural Gas Storage in Devonian Shales. SPE Formation Evaluation, **10**(02):109-113. http://dx.doi.org/10.2118/26632-PA

Lunatia,I and Lee. S.H. 2014. A dual-tube model for gas dynamics in fractured nanoporous shale formations. Journal of Fluid Mechanics. Vol (757): 943-971. http://dx.doi.org/10.1017/jfm.2014.519

Mahmoud, M.A. 2013. Development of a New Correlation of Gas Compressibility Factor (Z-Factor) for High Pressure Gas Reservoirs. Journal of Energy Resources and Technology, JERT-13-1048. http://dx.doi.org/10.1115/1.4025019

Maxwell,S. 2013. Beyond the SRV—The EPV Provides a More Accurate Determination of Reservoir Drainage in Shale Reservoirs. E&P. November Issue: 1-2

Michel G.G., Sigal, R.F., Civan, F., and Devegowda, D. 2011. Parametric Investigation of Shale Gas Production Considering Nano-Scale Pore Size Distribution, Formation Factor, and Non-Darcy Flow Mechanisms. Paper SPE 147438 presented at SPE Annual Technical Conference and Exhibition held in Denver, Colorado, 30 Oct – 2 November. http://dx.doi.org/10.2118/147438-MS

Mutyala, S., Fairbridge, C., Pare, J. R. J., et al. 2010. Microwave applications to oil sands and petroleum: A review. Fuel Process Technol , **91** (2), 127-135. http://dx.doi.org/10.1016/j.fuproc.2009.09.009

Newell, R. 2011, Shale Gas and the Outlook for U.S. Natural Gas Markets and Global Gas Resources. Report of U,S. Energy Information Administration. http://photos.state.gov/libraries/usoecd/19452/pdfs/DrNewell-EIA-Administrator-Shale-Gas-Presentation-June212011.pdf

Passey, Q., Bohacs, K., Esch, W., Klimentidis, R. and Sinha, S. 2010. From Oil-Prone Source Rock to Gas-Producing Shale Reservoir-Geologic and Petrophysical Characterization of Unconventional Shale Gas Reservoirs, International Oil and Gas Conference and Exhibition in China, 8-10 June. Society of Petroleum Engineers, SPE, Beijing, China

Raghavan, R., and Chin, L. Y. 2004. Productivity Changes in Reservoirs with Stress-Dependent Permeability. SPE Reservoir Evaluation & Engineering, **7**(04), 1–11. http://dx.doi.org/10.2118/88870-PA

Ramurthy, M., Barree, R. D., Kundert, D. P. et al. 2011. Surface-Area vs. Conductivity-Type Fracture Treatments in Shale Reservoirs. Paper SPE 140169 presented at the SPE Hydraulic Fracturing Technology Conference held in Woodlands, Texas, USA, 24-26 January. http://dx.doi.org/10.2118/140169-PA

Raymond, S., Aimene, E. Y., and Ouenes, A. 2015. Estimation of the Propped Volume Through the Geomechanical Modeling of Multiple Hydraulic Fractures Interacting with Natural Fractures. Paper SPE 176912 presented at the SPE Asia Pacific Unconventional Resources Conference and Exhibition, Brisbane, Australia, 9-11 November. http://dx.doi.org/10.2118/176912-MS

Roy, P., du Frane, W. L., and Walsh, S. D. C. 2015. Proppant Transport at the Fracture Scale: Simulation and Experiment. Paper presented at the 49th U.S. Rock Mechanics/Geomechanics Symposium, San Francisco, California, 28 June-1 July

Roy, S., Raju, R., Chuang, H.F., Cruden, B., and Meyyappan, M. 2003. Modeling gas flow through microchannels and nanopores. J. Appl. Phys. **93** (8): 4870–4879. http://dx.doi.org/10.1063/1.1559936

Sahai, R., Miskimins, J. L., and Olson, K. E. 2014. Laboratory Results of Proppant Transport in Complex Fracture Systems. Paper SPE 168579 presented at the SPE Hydraulic Fracturing Technology Conference, The Woodlands, Texas, 4-6 February. http://dx.doi.org/10.2118/168579-MS

Sahni, A., Kumar, M. and Knap, R.B. 2000. Electromagnetic heating methods for heavy oil reservoirs. Paper SPE 62550 presented at the SPE/AAPG Western Regional Meeting held in Long Beach, California, USA, 19-22 June. http://dx.doi.org/10.2118/62550-MS

Sakhaee-Pour, A., and Bryant, S. L. 2012. Gas Permeability of Shale. Paper SPE 146944-PA, SPE Reservoir Evaluation and Engineering, **15**(04):401–409. http://dx.doi.org/10.2118/146944-PA

Salmachi.A, Haghighi.M. Feasibility Study of Thermally Enhanced Gas Recovery of Coal Seam Gas Reservoirs Using Geothermal Resources. Energy & Fuels, 26(8): 5048-5059. http://dx.doi.org/10.1021/ef300598e

Sherman, F. 1969. The Transition From Continuum to Molecular Flow. Annual Review of Fluid Mechanics. Vol (1):317-340. http://dx.doi.org/10.1146/annurev.fl.01.010169.001533

Sigal, R. F., and Qin, B. 2008. Examination of the Importance of Self Diffusion In the Transportation of Gas In Shale Gas Reservoirs. Society of Petrophysicists and Well-Log Analysts. Petrophysics, 49(03):301-305. http://dx.doi.org/

Swami, V., Clarkson, C.R., and Settari, A. 2012. Non Darcy Flow in Shale Nanopores: Do We Have a Final Answer? Paper SPE 162665 presented at the Canadian Unconventional Resources Conference held in Calgary, Alberta, Canada, 30 October–1 November. http://dx.doi.org/10.2118/162665-MS

Symington, W.A., Olgaard, D.L., Otten, G.A., et al. 2006. ExxonMobil's Electrofrac Process for In Situ Oil Shale Conversion. Paper was presented at the 26th Oil Shale Symposium, held at the Colorado School of Mines in Golden, Colorado, October 16-18. http://dx.doi.org/ 10.1021/bk-2010-1032.ch010

Thoram, S. and Ehlig-Economides, C.A. 2011. Heat Transfer Applications for the Stimulated Reservoir Volume. Paper SPE 146975 presented at the SPE Annual Technical Conference and Exhibition, Denver, Colorado, USA. October 30–November 2. http://dx.doi.org/10.2118/146975-MS

U.S. Energy Information Administration. 2014. Newly released heat content data allow for state-to-state natural gas comparisons. Natural Gas Monthly. http://www.eia.gov/naturalgas/monthly/

Viswanathan, A. 2007. Viscosities of Natural Gases at High Pressures and High Temperatures. MS thesis, Texas AandM University, College Station, Texas

Wang, H. 2015. Numerical Modeling of Non-Planar Hydraulic Fracture Propagation in Brittle and Ductile Rocks using XFEM with Cohesive Zone Method. Journal of Petroleum Science and Engineering, Vol (135):127-140. http://dx.doi.org/10.1016/j.petrol.2015.08.010

Wang, H. 2016a. Poro-Elasto-Plastic Modeling of Complex Hydraulic Fracture Propagation: Simultaneous Multi-Fracturing and Producing Well Interference. Acta Mechanica, 227(2):507-525. http://dx.doi.org/ 10.1007/s00707-015-1455-7

Wang, H. 2016b. Numerical Investigation of Fracture Spacing and Sequencing Effects on Multiple Hydraulic Fracture Interference and Coalescence in Brittle and Ductile Reservoir Rocks. Engineering Fracture Mechanics, Vol (157): 107-124. http://dx.doi.org/10.1016/j.engfracmech.2016.02.025

Wang, H. 2016c. What Factors Control Shale Gas Production and Production Decline Trend in Fractured Systems: A Comprehensive Analysis and Investigation. SPE Journal (In Press). http://dx.doi.org/10.2118/179967-PA

Wang, H. and Marongiu-Porcu, M. 2015. Impact of Shale Gas Apparent Permeability on Production: Combined Effects of Non-Darcy Flow/Gas-Slippage, Desorption and Geomechanics. SPE-173196-PA. SPE Reservoir Evaluation & Engineering, **18** (04): 495-507. http://dx.doi.org/10.2118/173196-PA

Wang, H., Ajao, O., and Economides, M. J. 2014. Conceptual Study of Thermal Stimulation in Shale Gas Formations. Journal of Natural Gas Science and Engineering, Vol (21): 874-885. http://dx.doi.org/ 10.1016/j.jngse.2014.10.015

Wang, H., Marongiu-Porcu, M., and Economides, M. J. 2016. SPE-168600-PA. Poroelastic and Poroplastic Modeling of Hydraulic Fracturing in Brittle and Ductile Formations, Journal of SPE Production & Operations, **31**(01): 47–59. http://dx.doi.org/10.2118/168600-PA

Wang, H., Merry, H., Amorer, G. and Kong, B. 2015a. Enhance Hydraulic Fractured Coalbed Methane Recovery by Thermal Stimulation. Paper SPE 175927 presented at the SPE Unconventional Resources Conference to be held in Calgary, Alberta, Canada 20–22 Oct. http://dx.doi.org/10.2118/175927-MS

Wang, J., Liu,H., Wang, L., Zhang,H., Luo,H and Gao, Y. 2015b. Apparent permeability for gas transport in nanopores of organic shale reservoirs including multiple effects. International Journal of Coal Geology, Vol(152): 50-62. http://dx.doi.org/10.1016/j.coal.2015.10.004

Weng, X. 2014. Modeling of Complex Hydraulic Fractures in Naturally Fractured Formation. Journal of Unconventional Oil and Gas Resources, Vol. 9, pp. 114-135. http://dx.doi.org/ 10.1016/j.juogr.2014.07.001

Wong, R. C. K. and Yeung, K. C. 2006. Structural Integrity of Casing and Cement Annulus in a Thermal Well Under Steam Stimulation. J Can Pet Technol 45 (12): 6–9. http://dx.doi.org/10.2118/2005-104

Xu, B., Yuan, Y., and Wang, Z. 2011. Thermal Impact on Shale Deformation/Failure Behaviours - Laboratory Studies. Presented at the 45th U.S. Rock Mechanics/Geomechanics Symposium, San Francisco, 26–29 June.

Yahya, N, Kashif, M, Nasir, N, Akhtar, M. N, Yusof, N. M. 2012. Cobalt ferrite nanoparticles: An innovative approach for enhanced oil recovery application. Journal of Nano Research. Vol (17):115-126. http://dx.doi.org/ 10.4028/www.scientific.net/JNanoR.17.115

Ypma, T.J. 1995 Historical development of the Newton-Raphson method, SIAM Review, **37**(4), 531–551. http://dx.doi.org/10.1137/1037125

Yu, X., Rutledge, J., Leaney, S., and Maxwell, S. 2014. Discrete Fracture Network Generation from Microseismic Data using Moment-Tensor Constrained Hough Transforms. Paper SPE 168582 presented at the SPE Hydraulic Fracturing Technology Conference, The Woodlands, Texas, 4-6 February. http://dx.doi.org/10.2118/168582-MS

Yuan, Y., Xu, B., and Palmgren, C. 2013. Design of Caprock Integrity in Thermal Stimulation of Shallow Oil-Sands Reservoirs. J Can Pet Technol 52 (04): 266–278. http://dx.doi.org/10.2118/149371-PA

Yue, L.,Wang, H., Suai, H., and Nikolaou, M. 2015. Increasing Shale Gas Recovery through Thermal Stimulation: Analysis and an Experimental Study. Paper SPE 175070 presented at the SPE Annual Technical Conference and Exhibition held in Houston, Texas, USA, 28-30 Sep. http://dx.doi.org/10.2118/175070-MS

Zakhour, N., Sunwall, M., Benavidez, R., Hogarth, L., and Xu, J. 2015. Real-Time Use of Microseismic Monitoring for Horizontal Completion Optimization Across a Major Fault in the Eagle Ford Formation. Paper 173353 presented at the SPE Hydraulic Fracturing Technology Conference, The Woodlands, Texas 3–5 February. http://dx.doi.org/10.2118/173353-MS

Ziarani, A. S., and Aguilera, R. 2011. Knudsen's Permeability Correction for Tight Porous Media. Transport in Porous Media, **91**(1), 239–260. http://dx.doi.org/10.1007/s11242-011-9842-6

Zienkiewicz, O.C., and Taylor, R.L. 2005. *The Finite Element Method*, 5<sup>th</sup> edition, London: Elsevier Pte Ltd